\documentclass{birkjour}
\usepackage{amssymb}
\usepackage{bbold}
\usepackage{graphicx}
\usepackage{parskip}
\begin{document}

\title[Graphical Calculus for the Double Affine $Q$-Dependent Braid Group]
 {Graphical Calculus for the Double Affine\\ $Q$-Dependent Braid Group}

\author{Glen Burella}
\address{
Department of Mathematical Physics\\
National University of Ireland Maynooth\\
Maynooth, Co.\ Kildare\\
Ireland}
\email{glen.burella@nuim.ie}

\author{Paul Watts}
\address{
Department of Mathematical Physics\\
National University of Ireland Maynooth\\
Maynooth, Co.\ Kildare\\
Ireland}
\email{watts@thphys.nuim.ie}

\author{Vincent Pasquier}
\address{
Service de Physique Th\'eorique\\
CEA Saclay\\
91191 Gif-sur-Yvette\\
France}
\email{vincent.pasquier@cea.fr}

\author{Ji\v{r}\'i Vala}
\address{
Department of Mathematical Physics\\
National University of Ireland Maynooth\\
Maynooth, Co.\ Kildare\\
Ireland}  
\email{jiri.vala@nuim.ie}

\begin{abstract}
We define a double affine $Q$-dependent braid group.  This group is constructed by appending to the braid group a set of operators $Q_i$, before extending it to an affine $Q$-dependent braid group.  We show specifically that the elliptic braid group and the double affine Hecke algebra (DAHA) can be obtained as quotient groups. Complementing this we present a pictorial representation of the double affine $Q$-dependent braid group based on ribbons living in a toroid. We show that in this pictorial representation we can fully describe any DAHA.  Specifically, we graphically describe the parameter $q$ upon which this algebra is dependent and show that in this particular representation $q$ corresponds to a twist in the ribbon.
\end{abstract}

\maketitle

\section{Introduction}
\setcounter{equation}{0}
\renewcommand{\theequation}{\thesection.\arabic{equation}}

Representation theory is an essential tool in mathematical and physical research.
To this end, linear algebra, the theory of special
functions, arithmetic and related combinatorics are its usual
objectives.  A particularly potent example illustrating the power
of representation theory may be offered in the context of Hecke-type algebras \cite{IC1}.

In this paper we define a Hecke-type structure called
the double affine $Q$-dependent braid group and investigate its properties. 
Among its quotient groups is the double affine Hecke algebra (DAHA) 
which is of particular interest as its polynomial representations \cite{IC2} 
have close connections to
Macdonald and  Jack polynomials \cite{JL}.
Furthermore, we have seen how
some specific polynomials emerging from this algebra, when subject to
special wheel conditions, yield interesting $q$-deformed Laughlin and
Haldane-Rezayi wave functions \cite{MK,FJMM}.  These are believed to
be excellent candidates for describing quantum Hall effect ground
states; by adjusting the wheel condition parameters, one may fix
the filling fraction of these wavefunctions.  Other polynomials
directly obtained from the DAHA can, in a similar fashion, be used to
describe the ground states of $O(n)$ models \cite{KP}.

We provide an intuitive pictorial representation of a DAHA in this paper.
It is difficult to overestimate the power of graphical representations
in illustrating abstract concepts in pure mathematics.  Since the
emergence of the intuitive pictorial representation of the braid group
there has been a massive interest in its structure, greatly
advancing the field. For example, when Kauffman introduced diagrams
\cite{LK} to explain the Jones polynomial \cite{VFRJ} in the context
of Hecke algebras, the subject became more accessible and
widely-known.  

Before presenting our graphical representation which provides an interpretation of all DAHAs and their underlying parameters, we firstly establish the relation of DAHAs to other well known abstract algebraic structures. In particular we define and give readers a clear picture of the structure of a double affine $Q$-dependent braid group ($\mathcal{D}_{N}\{Q\}$). It is constructed by appending to the braid group a set of $N$ operators $\{Q_i\}=\{Q_1\ldots,Q_N\}$, before extending it to an affine $Q$-dependent braid group.  

\begin{figure}
\begin{center}
\includegraphics[scale=0.32]{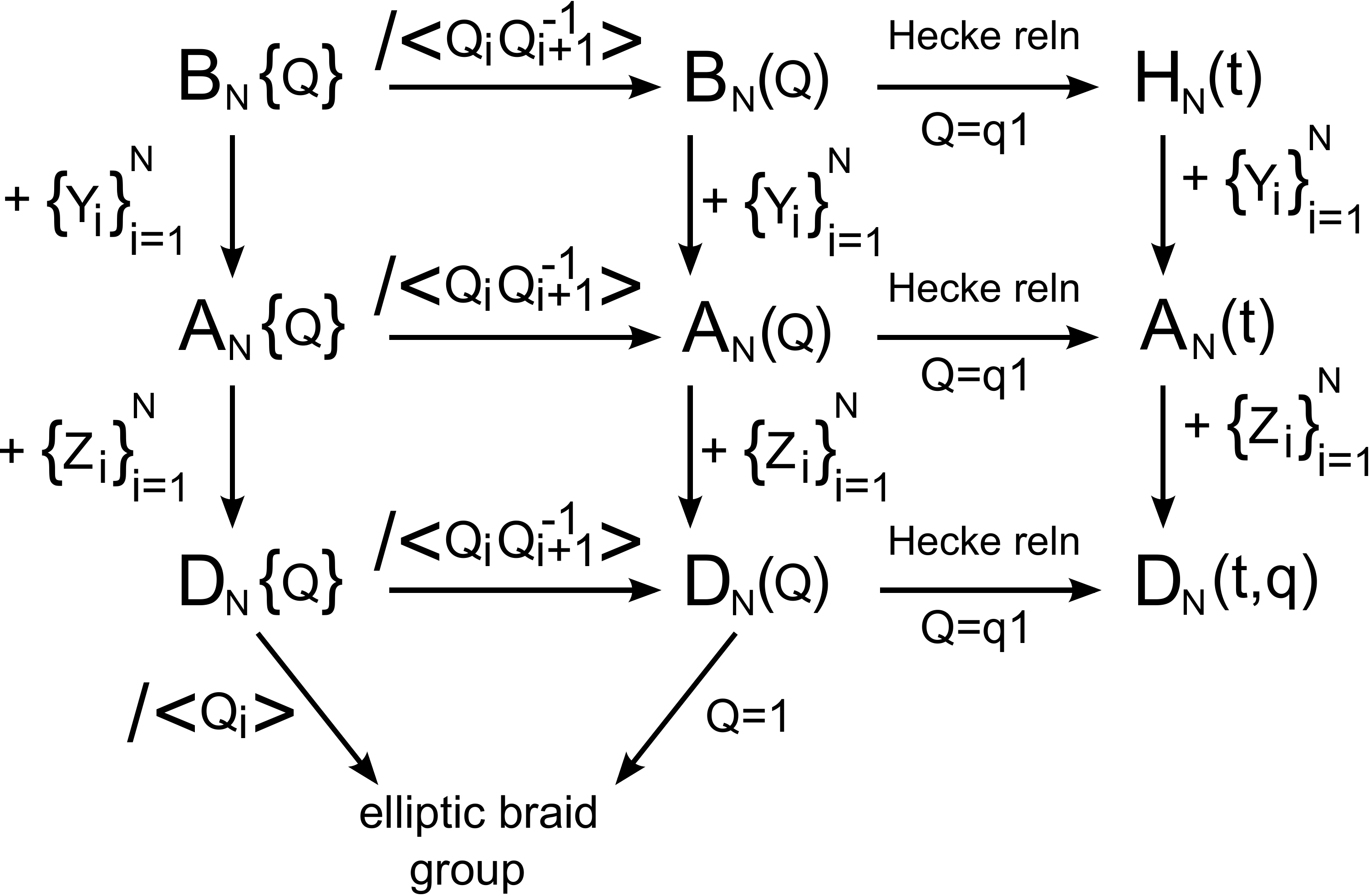} \label{F1}
\caption{Commutative diagram describing the relations of $\mathcal{D}_{N}\{Q\}$ with other algebraic structures whose primary element is a braid group.}
\end{center}
\end{figure}

Our interest in $\mathcal{D}_{N}\{Q\}$ stems from its pole position with respect to other algebraic structures whose primary element is a braid group. In fact, appending to the double affine braid group a set of operators $\{Q_i\}$ generalises the underlying braid group. It does so by turning braid group strands into ribbons and permitting $2\pi$ twists. The original braid group then corresponds to $\mathcal{B}_{N}\{Q\}/\langle Q_i\rangle$, where $\langle Q_i\rangle$ is freely generated by the operators $Q_i$. Thus the original braid group is in other words equivalent to  $\mathcal{B}_{N}\{Q\}$ where $Q=1$. Similarly the affine braid group corresponds to  $\mathcal{A}_{N}\{Q\}/\langle Q_i\rangle$. Naturally the elliptic braid group \cite{JB,GPS} is obtained from $\mathcal{D}_{N}\{Q\}$ by ignoring the twists or equivalently by contracting ribbons to strands, i.e. $\mathcal{D}_{N}\{Q\}/\langle Q_i\rangle$. In addition, taking the quotient
 $\mathcal{D}_{N}\{Q\}/\langle Q_iQ_{i+1}^{-1}\rangle$ is equivalent to considering twists on different ribbons as identical. Furthermore imposing the Hecke relation and setting $Q_i=q\mathbb{1}$, where $q \in \mathbb{C}$, we obtain the double affine Hecke algebra (of type A) \cite{IC1, BGHP}. These relations are illustrated in Figure \ref{F1}.  Note that $\mathcal{B}_{N}\{Q\}=\mathcal{B}_{N}(Q_1, Q_2,\ldots,Q_N)$ and $\mathcal{B}_{N}(Q)=\mathcal{B}_{N}\{Q\}/\langle Q_iQ_{i+1}^{-1}\rangle\simeq\mathcal{B}_{N}(Q, Q,\ldots,Q)$.
\smallskip

Complementing the algebraic description of a double affine $Q$-dependent braid group, we provide a pictorial representation.  The graphical calculus is based on ribbons within cubes, where opposite vertical faces of the cube are identified; a topologically equivalent presentation is given in terms of ribbons living inside a toroid.  We clearly illustrate all of the defining relations of $\mathcal{D}_{N}\{Q\}$ in our new cube-ribbon representation. It provides a concrete visual description of its structure, in particular we obtain a very straightforward interpretation of the
action of the generators $Q_i$ who create $2\pi$ twists in the ribbons. In the quotient group $\mathcal{D}_{N}\{Q\}/\langle Q_iQ_{i+1}^{-1}\rangle$, where we obtain the double affine Hecke algebra, we show that $q$ corresponds to the factor when replacing a ribbon with a
twist by one with no twist at all. Hence our cube-ribbon representation describes double affine Hecke algebras for all values of $q$. In $\mathcal{D}_{N}\{Q\}/\langle Q_i\rangle$ the ribbons are reduced to strands and twists are no longer possible, therefore our pictorial representation gives a toroidal description of the elliptic braid group. 

\bigskip

The layout of this paper is as follows: in Sections 1
through 3, we define the $Q$-dependent braid group and introduce the affine $Q$-dependent braid group.  We give their defining relations -- which depend on a set of operators $\{Q_i\}$ -- and pictorially represent their generators.

In Section 4 we present the complete construction of the double affine $Q$-dependent braid group. We outline our method of graphically representing this
group structure, which depends on the set of $\{Q_i\}$ and obtain the main result of this paper: that is we show that each generator $Q_i$ creates a twist in the ribbon. We also show that when $\{Q\}=1$ our cube-ribbon representation describes the elliptic braid group.

In Section 5 we indicate how to obtain the double affine Hecke algebra from $\mathcal{D}_{N}\{Q\}$. We highlight that our graphical calculus is valid for all DAHAs, with no restriction on the parameter $q$ upon which this algebra depends.\\
Finally we conclude with some discussion as to how this pictorial representation
could settle some unresolved issues, specifically regarding matrix and
tangle representations, and outline some related future work.

\section{The Braid Group}
\setcounter{equation}{0}
\renewcommand{\theequation}{\thesection.\arabic{equation}}

Throughout this paper we follow the general approach of \cite{MK, KP},
namely, we present all of the algebraic relations in terms of
multiplication rules for the elements of the algebra.  One could adopt
a much more rigourous approach via group quotients, etc.\ as in
\cite{IC1,IB}, but here we opt for this more ``physics'' approach.

\subsection{The $Q$-Dependent Braid Group $\mathcal{B}_{N}\{Q\}$}

We begin by reviewing the the braid group and its $Q$-dependent extension. These are essential to our construction of $\mathcal{D}_{N}\{Q\}$.  Similarly its well-established pictorial representation serves as a starting point for our cube-ribbon representation.

\bigskip

The $N$-strand braid group $\mathcal{B}_{N}$ is as follows \cite{EA}:
$\mathcal{B}_{N}$ is the group generated by the $N-1$ invertible
elements $\lbrace T_i| i=1,..,N-1 \rbrace$ satisfying the relations
\begin{eqnarray}
T_iT_j&=&T_{j}T_{i}\mbox{ for }|i-j| \geq 2, \label{T1}\\
T_{i}T_{i+1}T_{i}&=&T_{i+1}T_{i}T_{i+1}\mbox{ otherwise }\label{T2}.
\end{eqnarray}
(The second of the above is referred to as the {\it braid relation}.)

It is indeed well known that this algebraic description can be incorporated into a pictorial one by
defining $T_i$ and its inverse $T^{-1}_i$ to correspond to the
exchange of the $i^{\mathrm{th}}$ and $(i+1)^{\mathrm{th}}$ strands as
illustrated below:
\begin{center}
\includegraphics[scale=0.3]{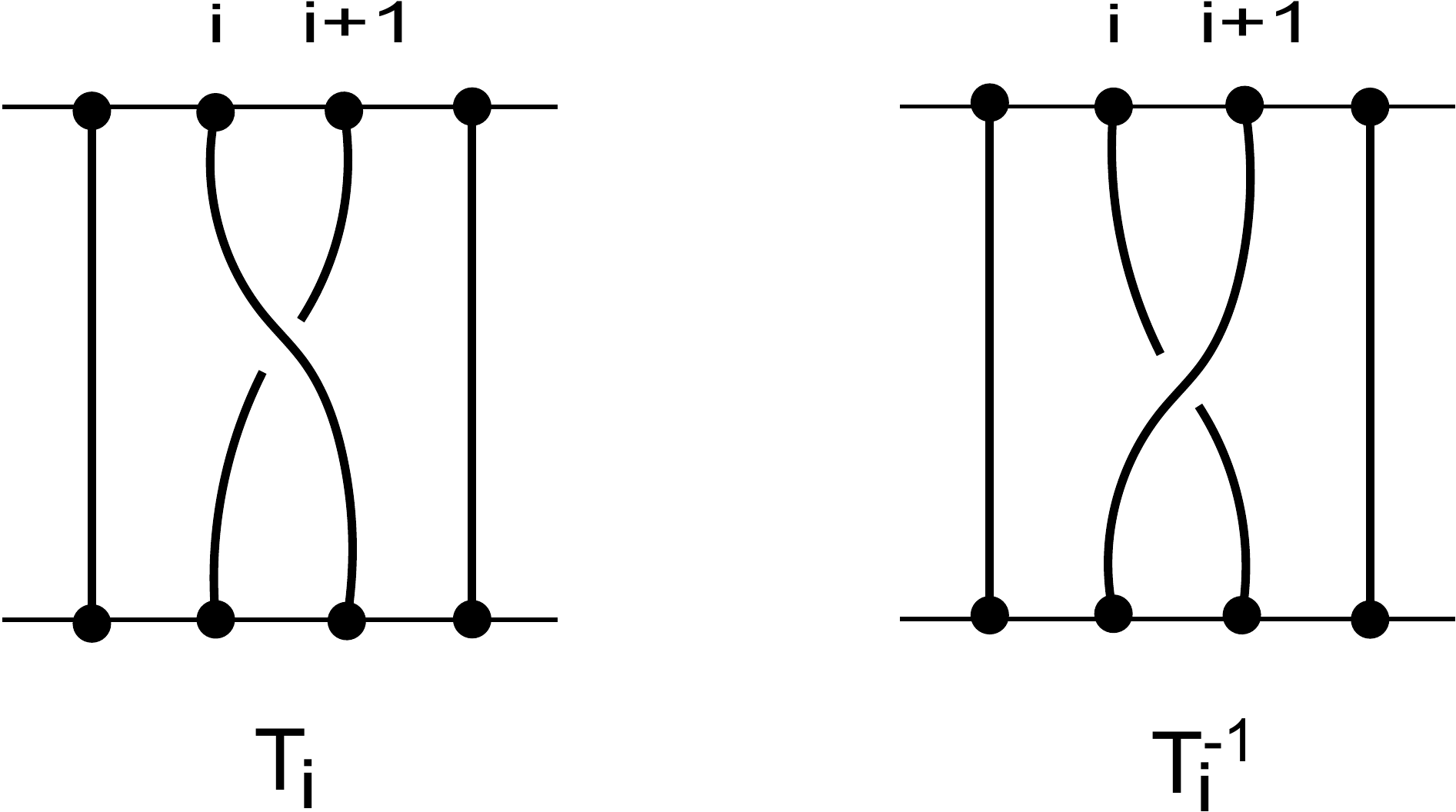}
\end{center}
Multiplication is then defined by stacking: AB is the braid obtained
by stacking A on top of B and gluing the bottom ends of the strands in A
to the top ends of those in B.  

\bigskip

We now define the $N$-strand $Q$-dependent braid group, $\mathcal{B}_{N}\{Q\}$, as follows:
$\mathcal{B}_{N}\{Q\}$ is the group generated by the invertible elements $\lbrace T_i| i=1,..,N-1 \rbrace$ satisfying (\ref{T1}) and (\ref{T2}), in addition to a set of commuting elements $\lbrace Q_i| i=1,..,N \rbrace$ satisfying the relations
\begin{eqnarray}
Q_iQ_j&=&Q_{j}Q_{i}\mbox{ for all }i,j,\label{T3}\\
T_{i}Q_{j}&=&Q_{j}T_{i}\mbox{ for } j<i, j>i+1,\label{T4}\\
T_{i}Q_{i}&=&Q_{i+1}T_{i}\mbox{ for }i=1,\ldots,N-1, \label{T5}\\
T_{i}Q_{i+1}&=&Q_{i}T_{i}\mbox{ for }i=1,\ldots,N-1. \label{T6}
\end{eqnarray}
These relations imply $T^2_iQ_j=Q_jT^2_i$ for all $i,j$, that is the $Q$s commute with all even powers of the $T$s but not with odd powers.

The above relations may be familiar to many readers.  They appear in the study of framed, or ribbon, braid groups, introduced in \cite{KO}. A more in-depth and mathematically rigourous treatment of framed braids or ribbon braids can be found in \cite{NW, TV}, among others. 
We use this well-known structure as a starting point for establishing the proper context for our treatment of DAHAs.

As it stands, only the trivial braids -- those whose strands go straight from top to bottom without crossing -- can represent the $Q$s in a way consistent with (\ref{T3})-(\ref{T6}).  We shall see later how to introduce nontrivial graphical representations for the $Q$s.

\section{Affine Braid Groups}
\setcounter{equation}{0}
\renewcommand{\theequation}{\thesection.\arabic{equation}}
\subsection{The Affine Braid Group $\mathcal{A}_N$}

The $Q$-dependent braid group $\mathcal{B}_{N}\{Q\}$  can be extended to an affine braid group $\mathcal{A}_N$ by appending to it $N$ invertible operators $Y_i$.  These satisfy the relations
\begin{eqnarray}
Y_iY_j &=& Y_jY_i\mbox{ for all }i,j,\label{Y1}\\
T_iY_j &=&Y_jT_i\mbox{ for }j\neq i, i+1,\label{Y2}\\
T_iY_{i+1}T_i &=& Y_i\mbox{ for }i=1,\ldots,N-1.\label{Y3}
\end{eqnarray}
The last of these relations implies that we need only one of the $Y_i$
(and all of the $T_i$) to generate the others.  For example, (\ref{Y3})
can be used to rewrite $Y_i$ for $i=2,\ldots,N$ as
\begin{eqnarray*}
Y_{i}&=&T^{-1}_{i-1}T^{-1}_{i-2}\ldots T^{-1}_1Y_1T^{-1}_1\ldots
T^{-1}_{i-2}T^{-1}_{i-1}.
\end{eqnarray*}

$\mathcal{A}_N$ is thus fully generated by $Y_1$ and the $T_i$. 

\bigskip

A more elementary presentation \cite{IC1,KP} is to write all the $Y_i$ in
terms of $T_i$ and an element $\sigma$ defined as
\begin{eqnarray}
\sigma&:=& T^{-1}_{N-1}T^{-1}_{N-2}\ldots T^{-1}_1Y_1.\label{sigma}
\end{eqnarray}
All of the $Y_i$ can now be written in terms of $\sigma$ and the $T_i$
using (\ref{Y3}):
\begin{eqnarray*}
\qquad Y_i &=& \left\{
\begin{array}{ll}
T_1T_2\ldots T_{N-1}\sigma&i=1\,,\\
T_i\ldots T_{N-1}\sigma T^{-1}_1\ldots T^{-1}_{i-1}&i=2,\ldots,N-1,\\
\sigma T^{-1}_1\ldots T^{-1}_{N-1}&i=N.
\end{array}
\right.
\end{eqnarray*}
The other defining relations for $\mathcal{A}_N$, (\ref{Y1}) and (\ref{Y2}),
may be rewritten in terms of $\sigma$ as
\begin{eqnarray*}
T_{i-1}\sigma = \sigma T_i,&& i=2,\ldots,N-1,\\
T_{N-1}\sigma^2=\sigma^2T_1.
\end{eqnarray*}

\smallskip

Also of interest is that the above relations imply that
$\sigma^NT_i=T_i\sigma^N$.  This tells us that $\sigma^N$ commutes
with all the $Y_i$, and thus $\sigma^N$ is central in
$\mathcal{A}_N$.  We could then label irreducible representations
of $\mathcal{A}_N$ with the eigenvalues of $\sigma^N$ if necessary.

\subsection{The Affine $Q$-Dependent Braid Group $\mathcal{A}_N\{Q\}$}

In a similar fashion to $\mathcal{B}_{N}\{Q\}$, we extend $\mathcal{A}_N$ to an affine $Q$-dependent braid group, $\mathcal{A}_N\{Q\}$, by defining how the set of elements $\lbrace Q_i| i=1,..,N \rbrace$ interact with the affine generators $Y_i$.
\smallskip

Therefore in addition to all of the defining relations of $\mathcal{A}_N$, the generators of $\mathcal{A}_N\{Q\}$ must also satisfy
\begin{eqnarray}
Y_{i}Q_{j}&=&Q_{j}Y_{i}\mbox{ for all }i,j.\label{Y4}
\end{eqnarray}

Using the definition of $\sigma$, (\ref{sigma}), one can rewrite (\ref{Y4}), to obtain $\mathcal{A}_N\{Q\}$ purely in terms of $T_i$, $\sigma$ and $Q_i$:
\begin{eqnarray*}
\sigma Q_i&=&Q_{i-1}\sigma\mbox{ for }i=2,\ldots,N,\\
\sigma Q_1&=&Q_N\sigma.
\end{eqnarray*}

These relations also imply that $\sigma^NQ_i=Q_i\sigma^N$.  Having fully described our definition of an affine $Q$-dependent braid group, $\mathcal{A}_N\{Q\}$, we now incorporate its algebraic structure into an intuitive graphical one.

\subsection{Pictorially Representing $\mathcal{A}_N\{Q\}$}

We have already seen that in the pictorial representation of the braid
group $\mathcal{B}_{N}$, the braiding of the strands takes place in
the strip in a strict top-to-bottom direction.  Now we turn the strip
into a cylinder by identifying the left and right edges; to highlight
this point, we represent these edges with dashed lines.  This means
that we can now braid in a left-to-right (or vice versa) fashion by
wrapping strands around the cylinder.  This application of cyclic
boundary conditions is what gives us a pictorial representation for
the affine $Q$-dependent braid group $\mathcal{A}_N\{Q\}$.  (The braid group
generators $T_i$ still braid top-to-bottom as they did before we
identified the sides.)

To illustrate this, we define the pictorial representations of the $\mathcal{A}_N\{Q\}$
generator $Y_i$ and its inverse $Y^{-1}_i$ as follows:
\begin{center}
\includegraphics[scale=0.28]{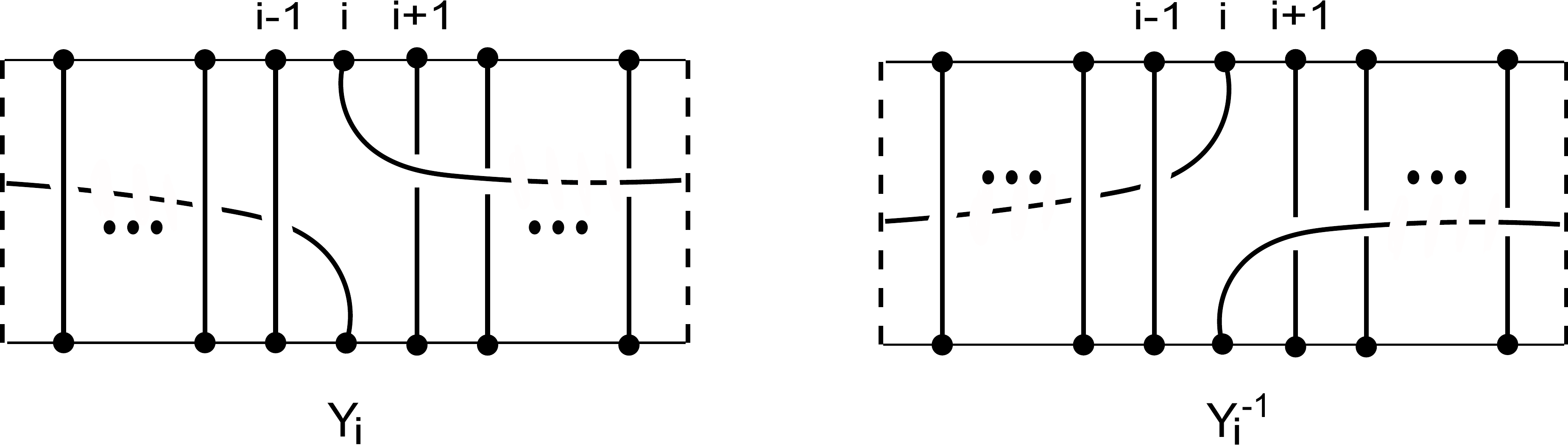}
\end{center}

So we see that $Y_i$ takes the strand starting at point $i$ on the top
edge and takes it to the same point on the bottom edge and leaves all
other strands untouched, and does so such that it goes over all
strands to the right ($i+1,\ldots,N$) and under all strands to the
left ($1,\ldots,i-1$).  For example, in the $N=3$ case, $Y_1$ is given
by either of the two pictures below:
\begin{center}
\includegraphics[scale=0.28]{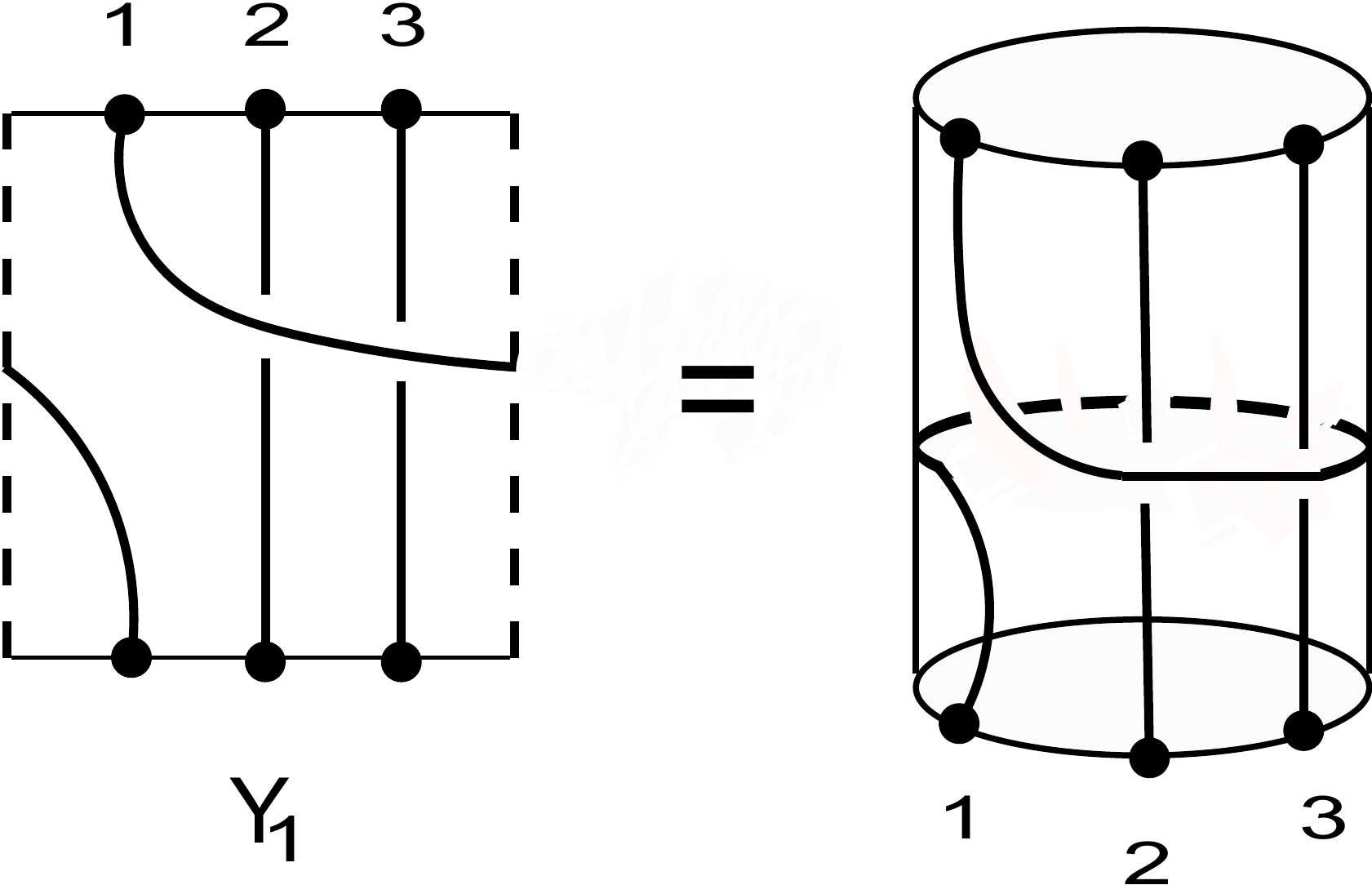}
\end{center}

Multiplication is now defined by stacking cylinders on top of one
another, and so given $Y_1$ and the $T_i$, we can construct all other
$Y_i$ via (\ref{Y3}).  Looking at the $N=3$ case again, we can now
construct $Y_2=T^{-1}_1Y_1T^{-1}_1$ and see that our pictorial
representation is consistent:
\begin{center}
\includegraphics[scale=0.28]{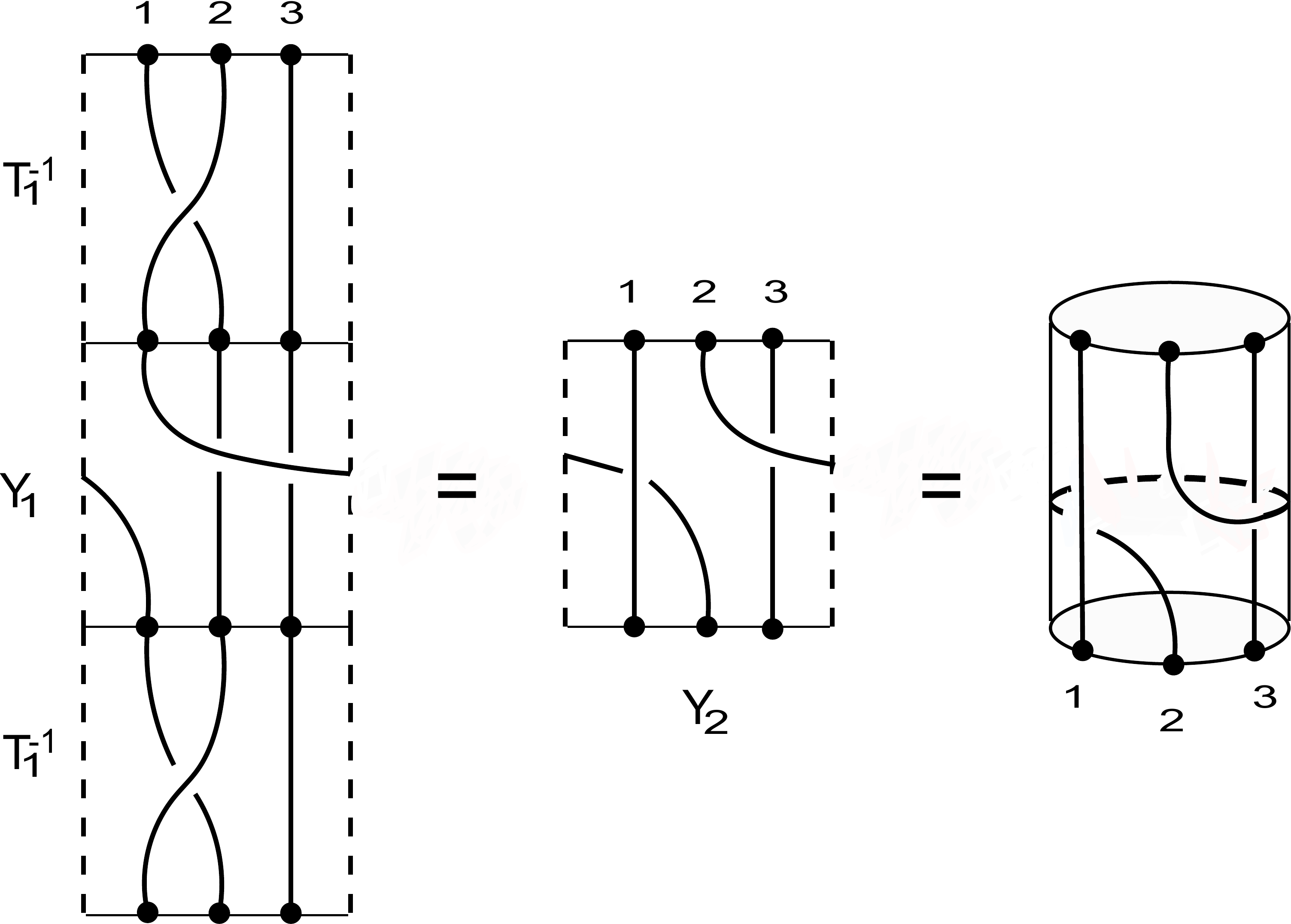}
\end{center}

Recall, from (\ref{sigma}), that $\sigma$ was defined in terms of
$Y_1$: $\sigma = T^{-1}_{N-1}T^{-1}_{N-2}....T^{-1}_1Y_1$.  Therefore,
for $N=3$, we have $\sigma=T^{-1}_2T^{-1}_1Y_1$, which looks like
\begin{center}
\includegraphics[scale=0.30]{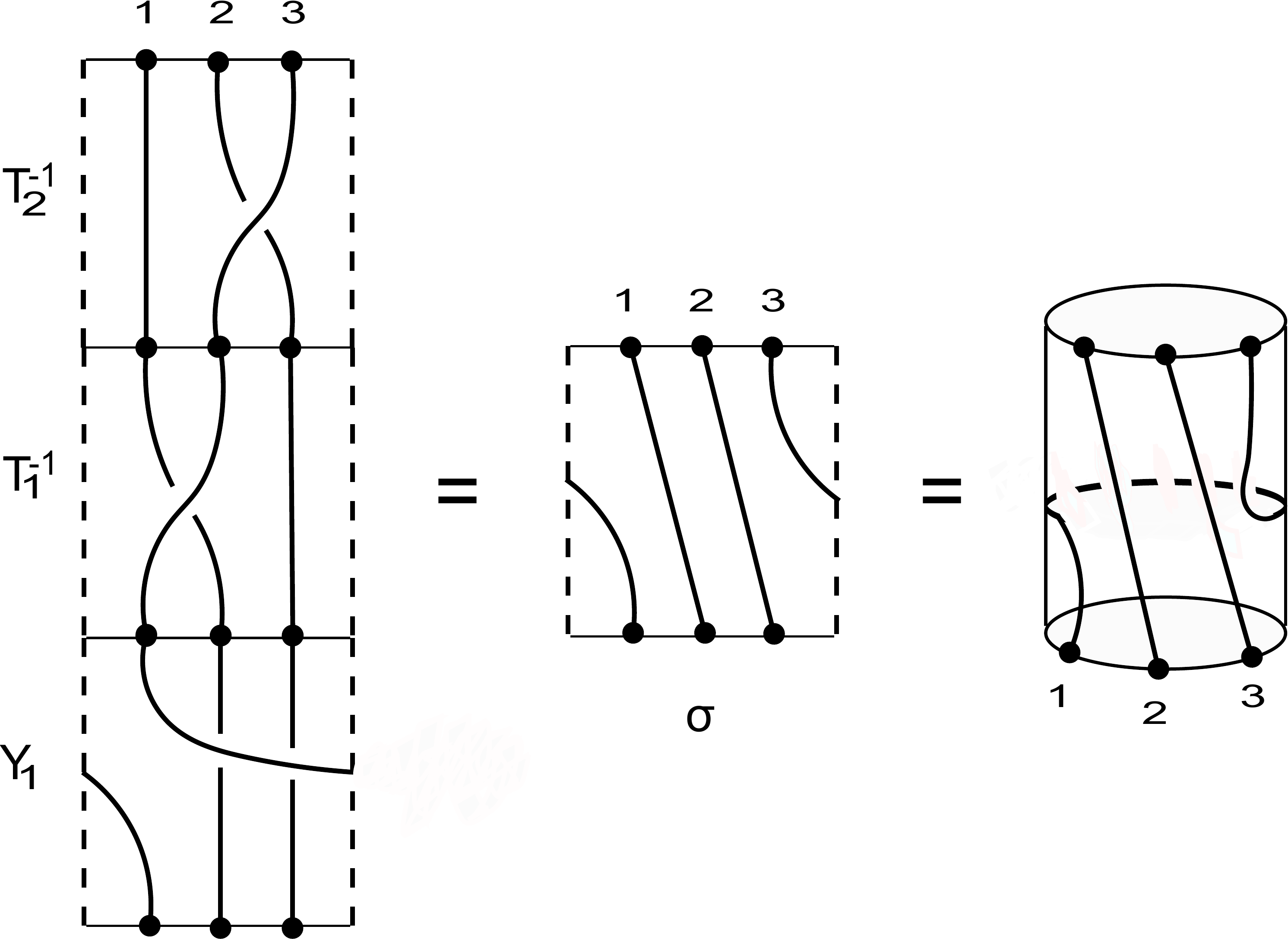}
\end{center}
$\sigma$ has the same general form for all $N$, namely, it acts as a
kind of raising operator on the indices by taking point $i$ on the top
to point $i+1$ on the bottom (with the cylindrical topology
identifying point $N+1$ with $1$).  Therefore, we take this to be the
pictorial definition of $\sigma$, and so together with the cylinders
representing the $T_i$, all of the defining relations of the $\mathcal{A}_N\{Q\}$
follow suit.

\smallskip

At this point we have a complete pictorial representation for the $Y$s.  However, the $Q$s are still only representable by trivial braids.  Despite this we can extend $\mathcal{A}_N\{Q\}$ to a double affine $Q$-dependent braid group by incorporating a whole new set of generators and their graphical representations, as we will now show.

\section{Double Affine Braid Groups}
\setcounter{equation}{0}
\renewcommand{\theequation}{\thesection.\arabic{equation}}
\subsection{The Double Affine $Q$-Dependent Braid Group $\mathcal{D}_N\{Q\}$}

We can extend $\mathcal{A}_N\{Q\}$ to a double affine $Q$-dependent braid
group $\mathcal{D}_N\{Q\}$ \cite{IC1,BGHP} by
introducing a further $N$ invertible generators $Z_i$ satisfying the
relations
\begin{eqnarray}
Z_iZ_j &=& Z_jZ_i\mbox{ for all }i,j,\label{Z1}\\
T_iZ_j &=&Z_jT_i\mbox{ for }j\neq i, i+1,\label{Z2}\\
T_iZ_{i+1}T_i &=& Z_i\mbox{ for }i=1,\ldots,N-1,\label{Z3}
\end{eqnarray}
together with the set of elements $\lbrace Q_i| i=1,..,N \rbrace$ which commute with all the $Z_i$ and appear explicitly in relations intertwining the $Y_i$ and the $Z_i$ \cite{IC1}:
\begin{eqnarray}
Z_iQ_j&=&Q_jZ_i\mbox{ for all }i,j,\label{Z4}\\
Y_1Z_2Y^{-1}_1Z^{-1}_2 &=& T^2_1,\label{YZT}\\
Y_i\left(\displaystyle\prod_{j=1}^N Z_j\right)
 &=&Q_i\left(\displaystyle\prod_{j=1}^N Z_j\right)Y_i,\label{YZ}\\
Z_i\left(\displaystyle\prod_{j=1}^NY_j\right) &=&
Q^{-1}_i\left(\displaystyle\prod_{j=1}^N Y_j\right)Z_i.\label{ZY}
\end{eqnarray}

We can choose to eliminate the $Y_i$ in favour of the cyclic operator
$\sigma$, and then (\ref{YZT}) and (\ref{YZ}) can be rewritten as
\begin{eqnarray*}
Z_{i-1}\sigma = \sigma Z_i,&& i=2,\ldots,N,\\
Z_N\sigma=Q^{-1}_N \sigma Z_1.
\end{eqnarray*}

Using the above relations, one can quickly see that
\begin{eqnarray}
Z_i\sigma^N& =& Q^{-1}_i\sigma^NZ_i,\label{Z-sigma}
\end{eqnarray}
and this, in addition to the identity $\prod_{j=1}^NY_j=\sigma^N$
\cite{IC1} (a proof of which we include in Appendix A for the
interested reader) gives us (\ref{ZY}).  Therefore it is not independent of
the other relations.

\bigskip

To summarise, we define a $\mathcal{D}_N\{Q\}$ to be the group generated by $T_i$,
$Y_i$, $Z_i$ and $Q_i$ which satisfy equations (\ref{T1})-(\ref{T6}),
(\ref{Y1})-(\ref{Y3}) alongside (\ref{Y4}) and (\ref{Z1})-(\ref{YZ}).  We shall see shortly
that the appearance of the operators $Q_i$ in the last of these defining
relations will strongly influence our choice of pictorial
representation for $\mathcal{D}_N\{Q\}$.

\subsection{Graphical Representation of $\mathcal{D}_N\{Q\}$}

Recall that we extended the standard pictorial representation of the
braid group to that of an $\mathcal{A}_N\{Q\}$ by identifying the two vertical edges
and defining the action of the $Y_i$ generators on the strands as
wrapping around the resulting cylinder.  We would now like to extend
this $\mathcal{A}_N\{Q\}$ representation to one for a $\mathcal{D}_N\{Q\}$ by somehow incorporating the new generators $Z_i$ into the picture.

Our method for doing so is motivated by the $\mathcal{A}_N\{Q\}$ construction: the
braid group generators do not wind strands at all; they simply connect
points on the top edge to ones on the bottom.  The $Y_i$ generators,
however, do wind the strands ``perpendicular'' to the $T_i$, namely,
left-to-right (or vice versa) instead of top-to-bottom.

The new $Z_i$ generators have exactly the same relations between
themselves and the $T$s as the $Y$s do, so this suggests that we need
a {\em third} direction.  This suggests that
instead of a strip whose two vertical sides are identified, we now use
a {\em cube} whose opposite vertical faces are identified.  So the
left and right faces of the cube are identified with the $Y_i$
operators taking strands through them, while the front and back faces
are identified with the $Z_i$ generators taking strands through them.

To see this, first consider drawing each braid group generator $T_i$
in a cube.  The braiding now takes place within the cube from top to
bottom:
\begin{center}
\includegraphics[scale=0.28]{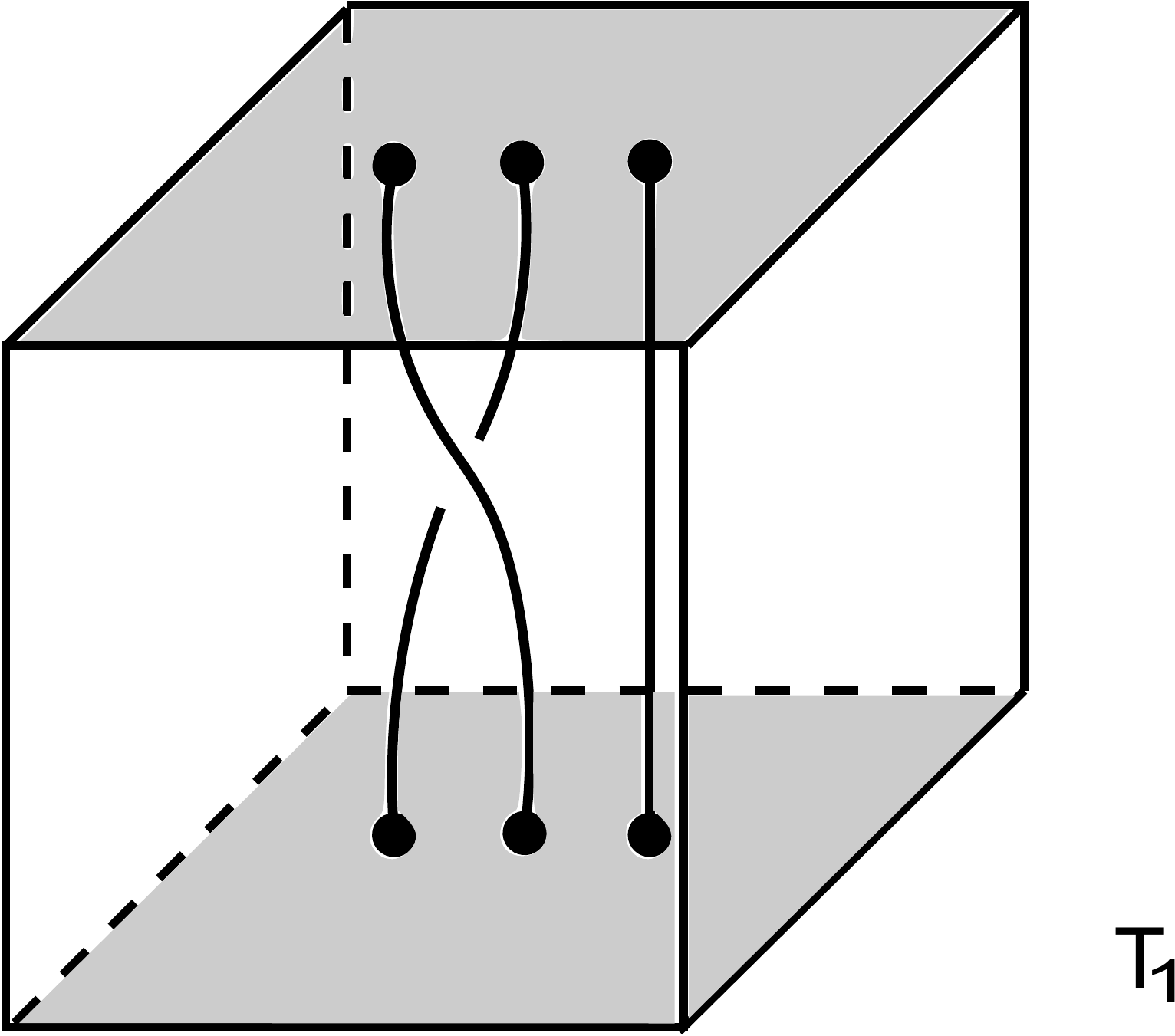}
\end{center}
Multiplication is defined in the usual way, by stacking one cube onto another.

This representation is essentially the same as that for the elliptic
braid group on a torus \cite{JB,GPS}, which is generated by $T_i$,
$Y_i$ and $Z_i$ but requires all the $Q_i$ to be
unity.  In Section \ref{4.3}, we show that the $Q_i$ are indeed $\mathbb{1}$ for our
representation, as expected.  This is not a surprising result though as the elliptic 
braid group is simply $\mathcal{D}_N\{Q\}$/$\langle Q_i\rangle$.  However, using three-dimensional cubes
rather than a two-dimensional torus will allow us to generalise to
values of $Q_i$ other than unity, as we illustrate in Section \ref{4.3.2}.

\bigskip

Recall that the affine $Q$-dependent braid group generators $Y_i$ identified the
left and right sides with each other to give braiding on a cylinder.
In the cube representation, we identify the left and right faces of
the cube with each other.  In the figure below, the turquoise arrows
traverse the coloured blue planes and wrap the strand around the cube
from one to the other.
\begin{center}
\includegraphics[scale=0.28]{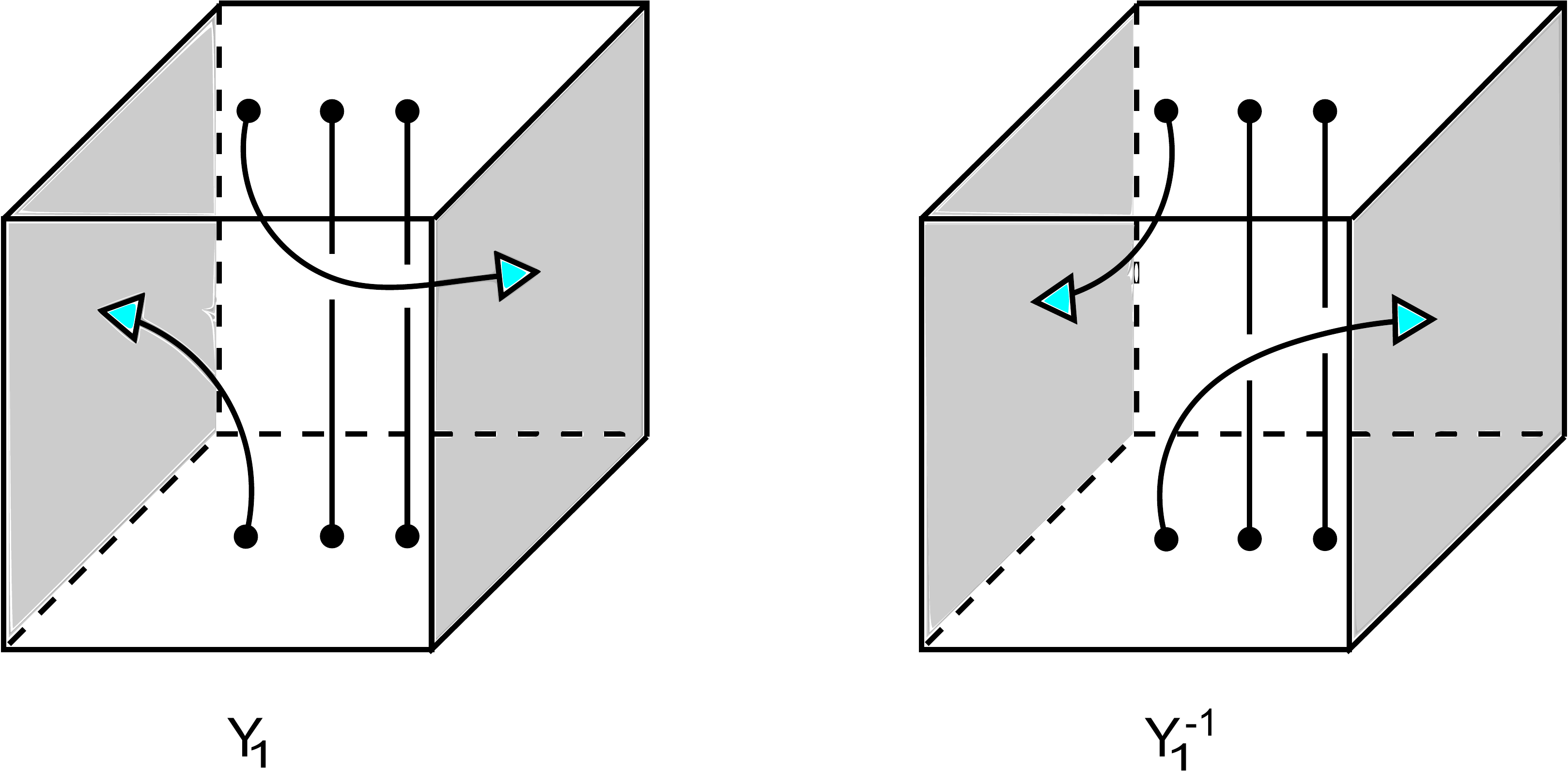}
\end{center}
The additional $\mathcal{D}_N\{Q\}$ generators $Z_i$ identify the front face of the
cube with its back face.  In the figure below, we use red tips to
indicate that the strand passes out through the coloured front face of
the cube, then wraps around until it meets the strand that passes out
the back face.  More specifically, for the $N=3$ case we define $Z_1$
(and its inverse) as
\begin{center}
\includegraphics[scale=0.28]{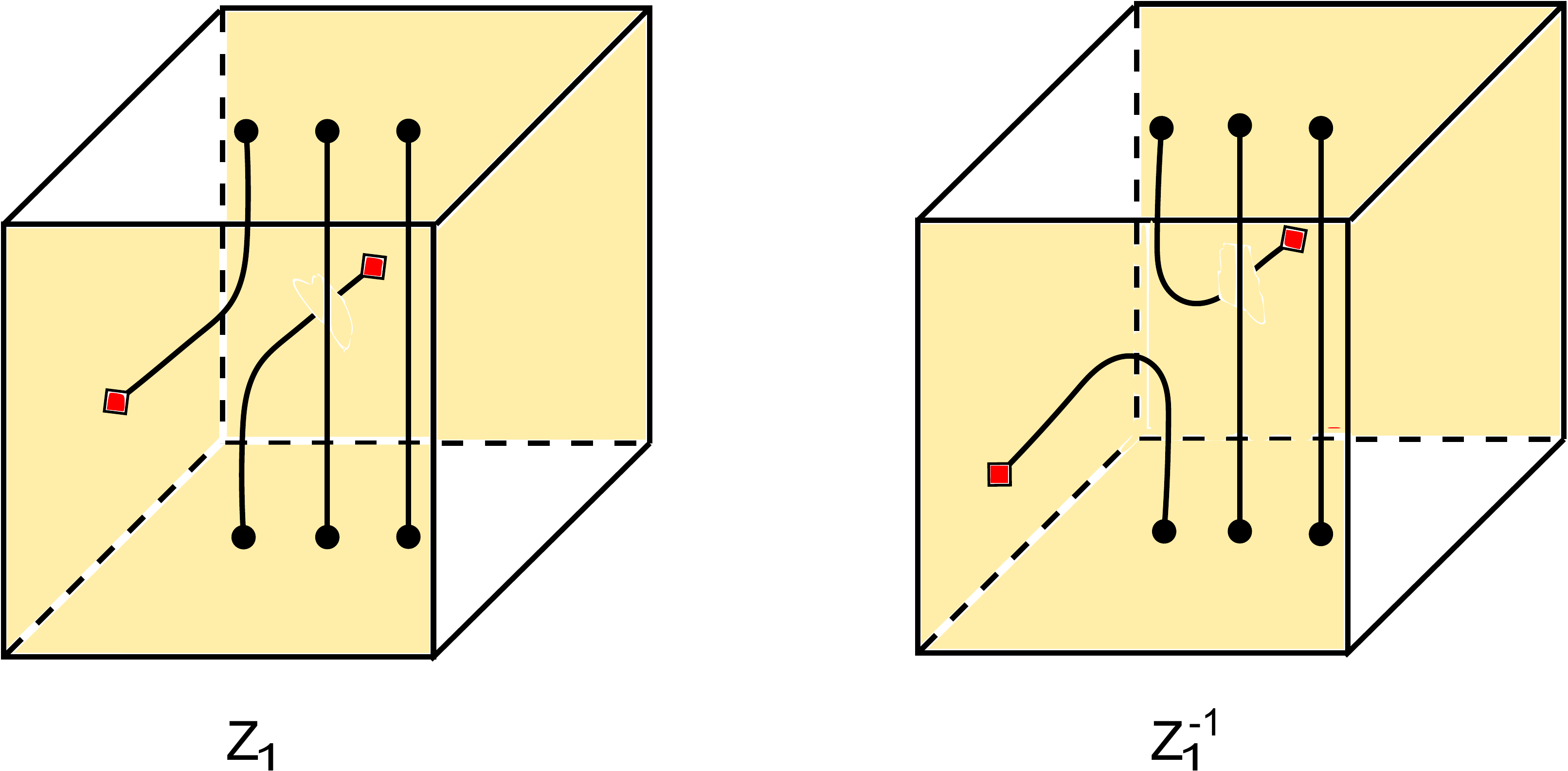}
\end{center}
Having defined $Z_1$, we can now obtain all of the other $Z_i$ for
$i=2,\ldots,N$ using $T_iZ_{i+1}T_i=Z_i$.  So, for example,
$Z_2=T^{-1}_1Z_1T^{-1}_1$:
\begin{center}
\includegraphics[scale=0.28]{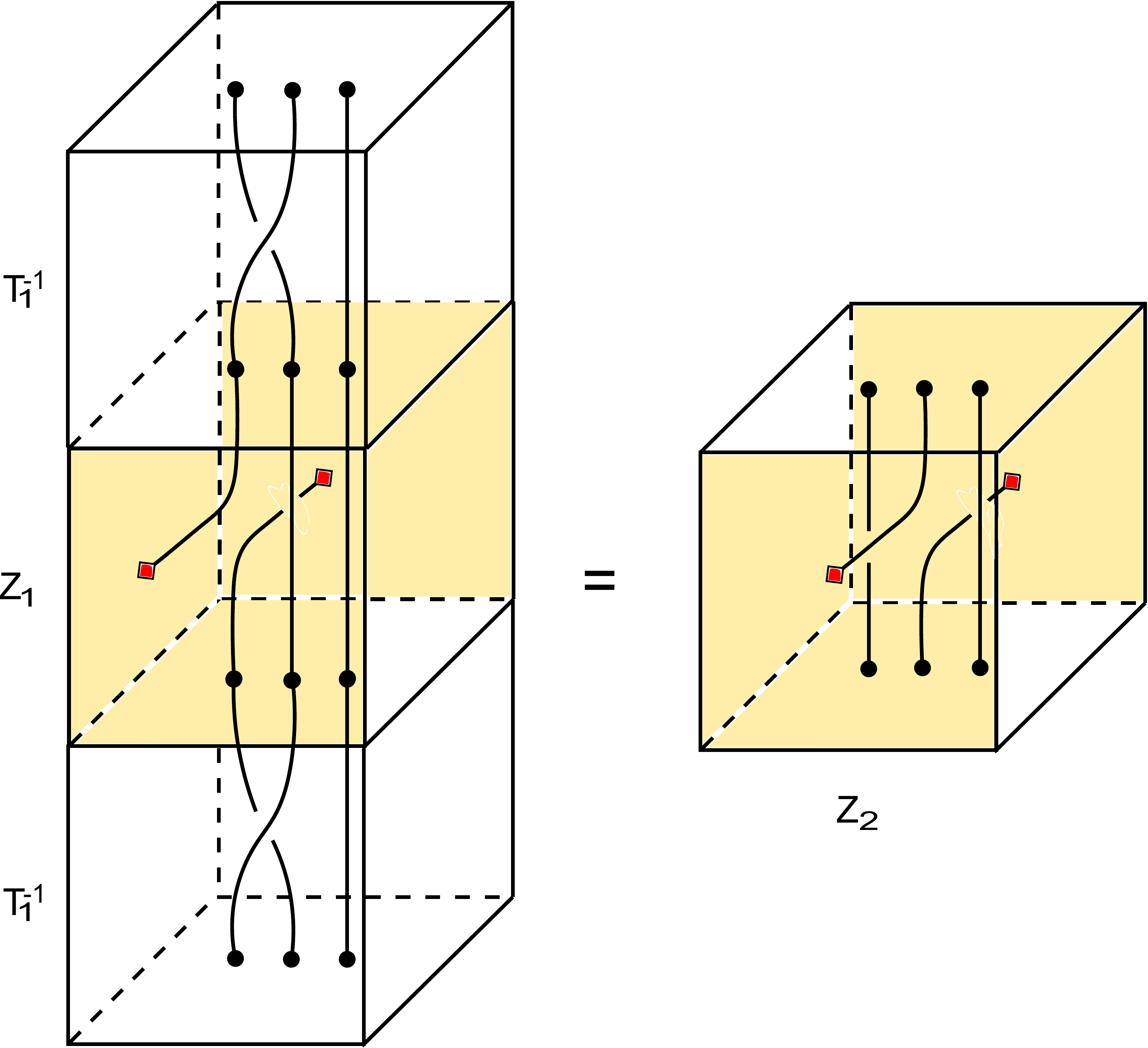}
\end{center}
One may proceed in this manner to construct $Z_i$ for any $i$, and we
see that its action is to take the $i^{\mathrm{th}}$ point on the top
face, bring it out the front face of the cube, wrap around to come in
the back face, and connect to the $i^{\mathrm{th}}$ point on the
bottom, with all other strands simply going straight from top to
bottom.

At this point, we note that our cube is topologically equivalent to a
hollowed-out toroid: identification of the opposing sides of any
horizontal slice of the cube gives a $2$-torus, and the region between
the top and bottom faces -- a time interval $I$ if we view our strands
as worldlines -- gives the thickness.  Thus, each of our generators is
represented as $N$ strands within the toroid $S^1\times S^1\times I$.

To illustrate this further, define two angles, $\theta$ and $\varphi$.
We let $\theta$ be the direction in which the $Y_i$ generators wrap
around the toroid and $\varphi$ is the direction the $Z_i$ wrap around
the toroid.  So, in effect,
the $\mathcal{A}_N\{Q\}$ generators $Y_i$ encircle the torus within the toroid whereas
the additional $\mathcal{D}_N\{Q\}$ generators $Z_i$ encircle the empty space bounded
by the toroid, as illustrated below:
\begin{center}
\includegraphics[scale=0.25]{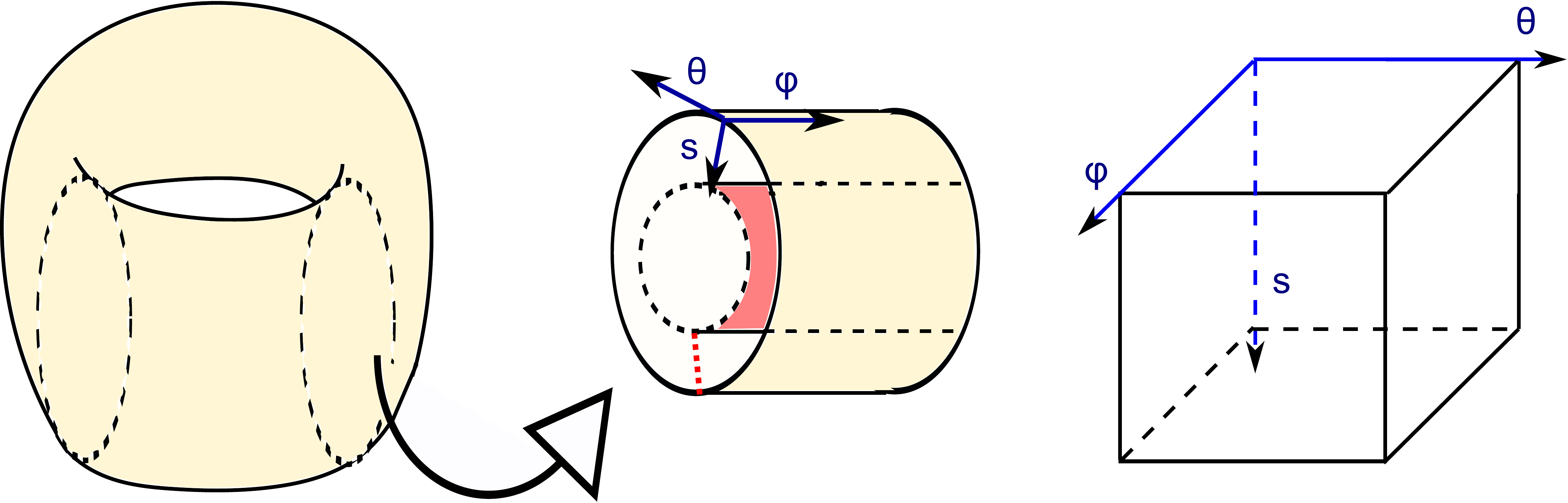}
\end{center}
\vspace{1em}
where $s\in I$ is the time parameter.  In this toroidal
representation, one can now clearly see the distinct directions in
which the different generators wrap.
\begin{center}
\includegraphics[scale=0.27]{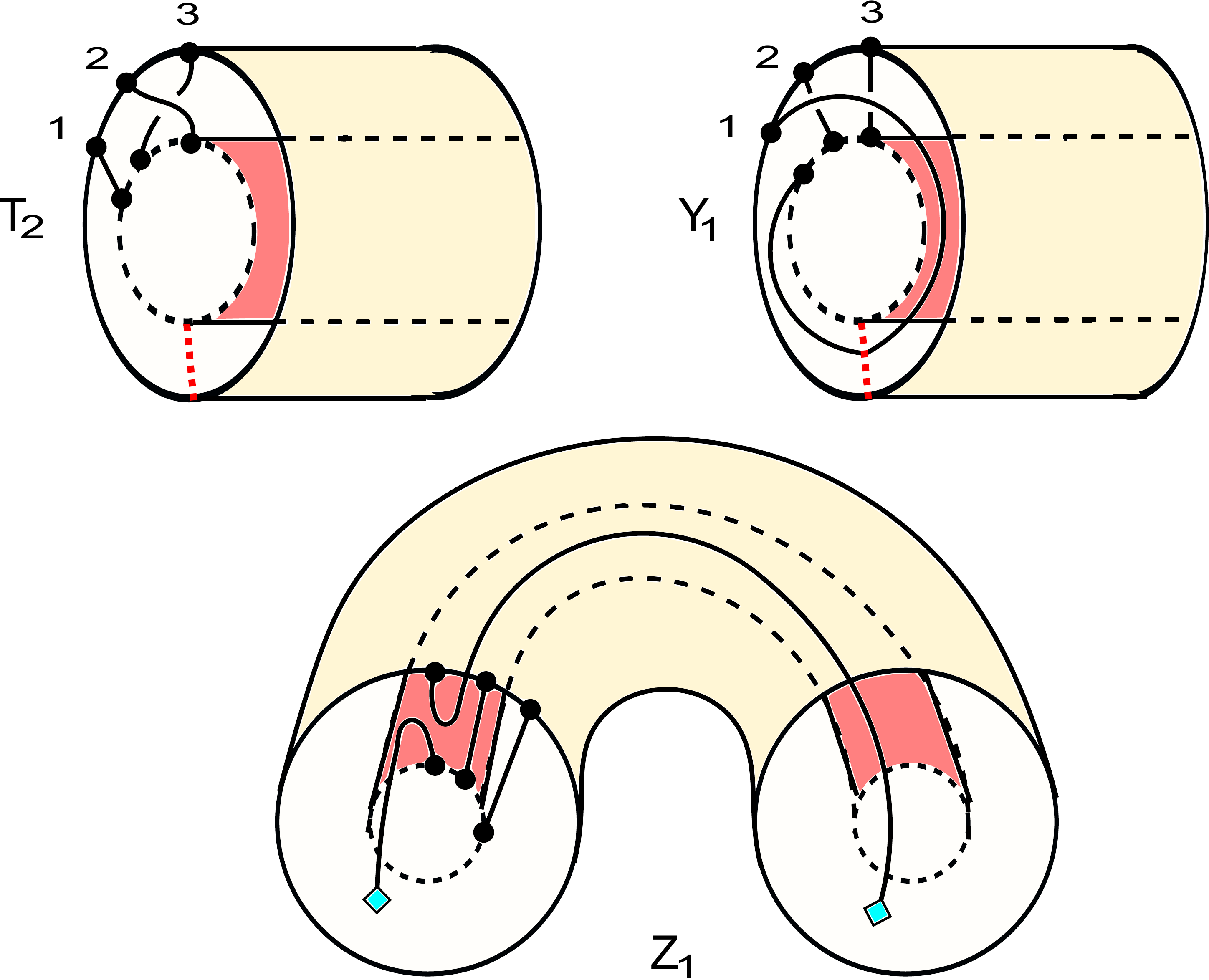}
\end{center}

\begin{center}\begin{figure}[h!]
\includegraphics[scale=0.26]{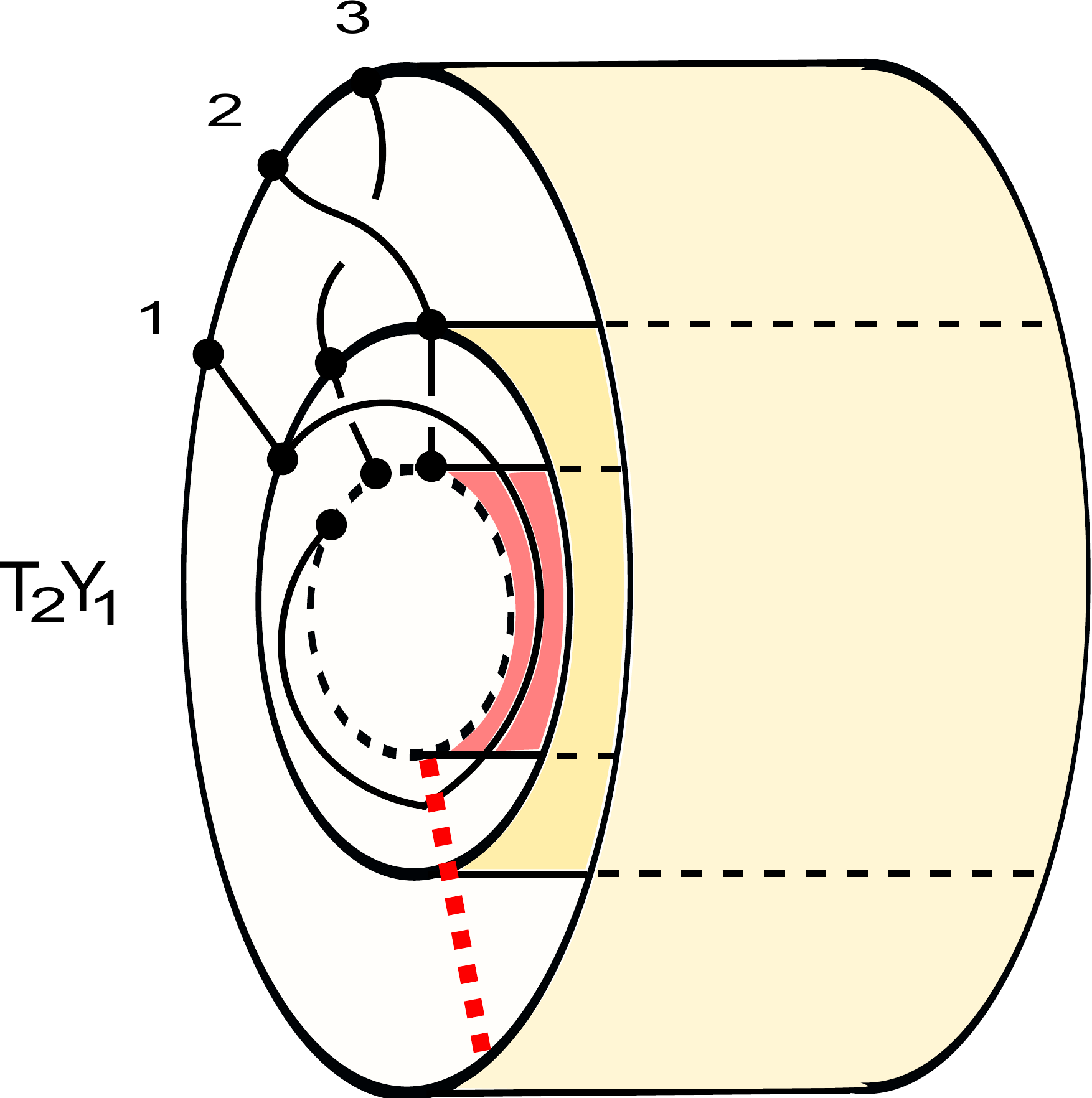}
\caption{Toroidal representation of the product $T_2Y_1.$}
\label{torusmult}
\end{figure}\end{center}

Multiplication is defined by stuffing toroids inside each other: this
is done such that the points on the inner boundary of the first (in
order of multiplication) generator correspond to the points on the
outer boundary of the second generator.  In Figure \ref{torusmult} we
illustrate the product $T_2Y_1$: we stuff $Y_1$ into $T_2$ such that
the numbered points on the outer boundary of $Y_1$ correspond to the
points on the inner boundary of $T_2$.

\subsection{Graphical Representation of the action of $Q_i$}\label{4.3}
\subsubsection{The case $Q_i=\mathbb{1}$}\label{4.3.1}

\begin{center}\begin{figure}[h!]
\includegraphics[scale=0.20]{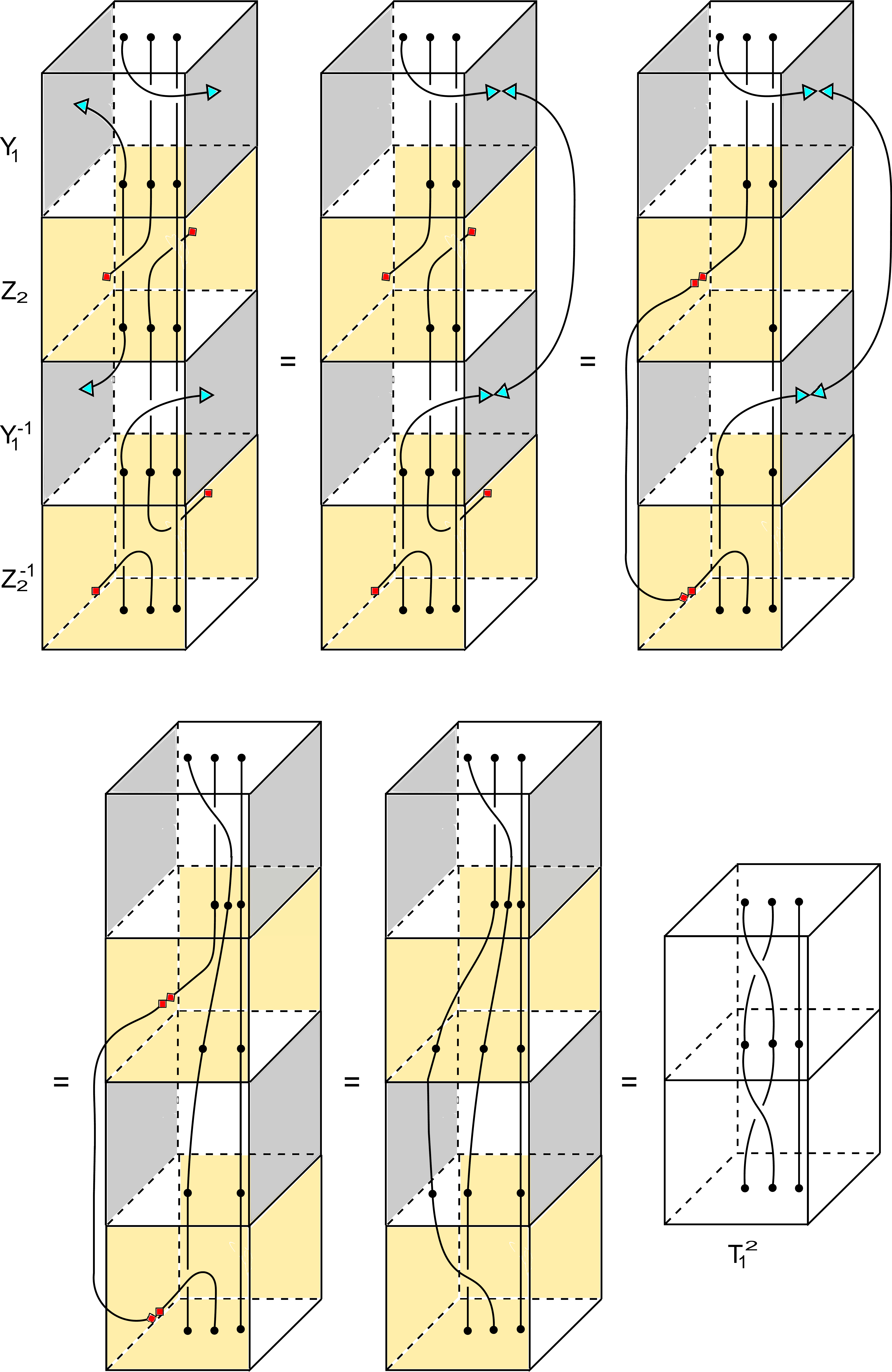}
\caption{Step-by-step verification of the relation $Y_1Z_2Y_1^{-1}Z_2^{-1}=T_1^2$ in the cube representation.  }
\label{T_1^2=Y_1Z_2Y_1Z_2}
\end{figure}\end{center}

We must confirm that our cubic/toroidal representation works for all
the $\mathcal{D}_N\{Q\}$ axioms.  We start by verifying (\ref{YZT}),
i.e.\ $Y_1Z_2Y^{-1}_1Z^{-1}_2 = T^2_1$.  From Figure
\ref{T_1^2=Y_1Z_2Y_1Z_2}, we see that this
is satisfied by our cube representation.

\begin{center}\begin{figure}[h!]
\includegraphics[scale=0.28]{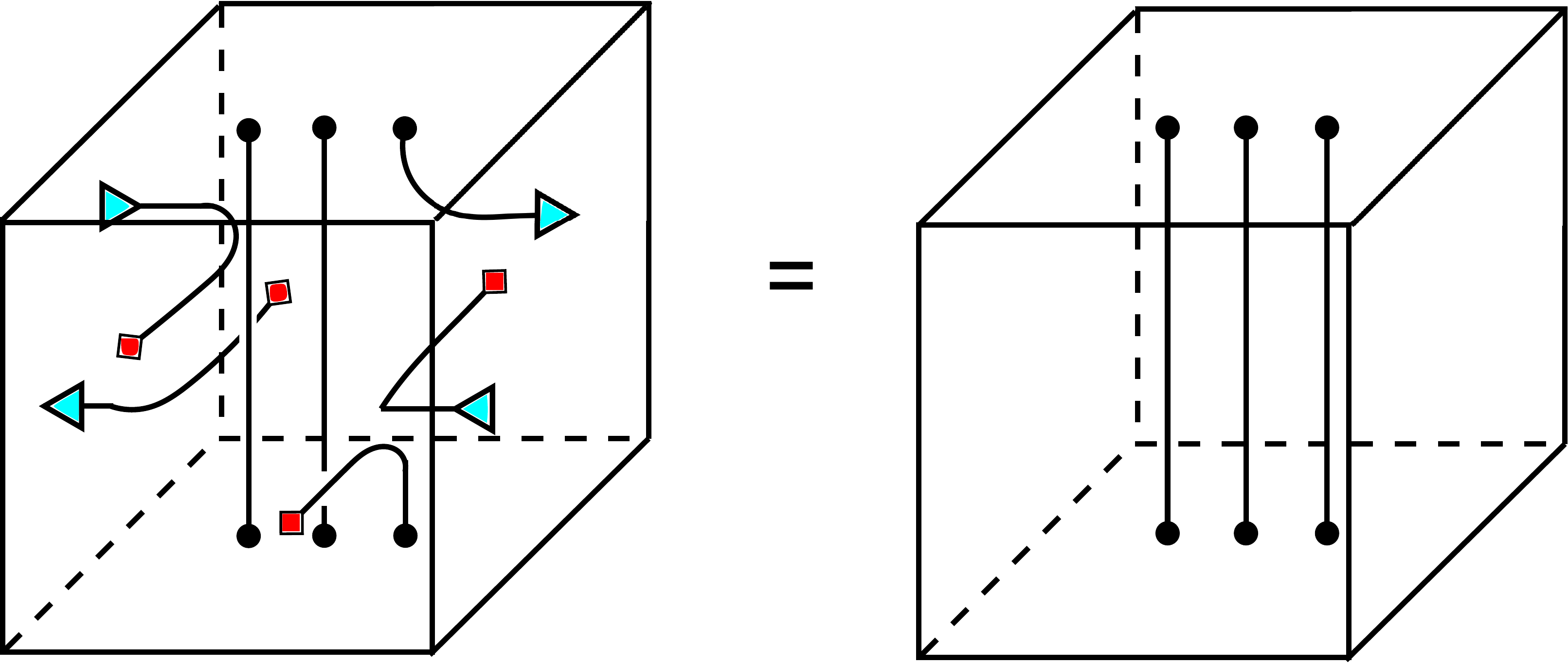}
\caption{Pictorial representation of
$Q_3=\sigma Z_1\sigma^{-1}Z^{-1}_{3}$.  Pulling all 
strands tight yields the identity.}
\label{Q_3}
\end{figure}\end{center}

(\ref{YZ}) and (\ref{ZY}) must also hold in our representation, of
course.  These are the relations that depend explicitly on the
elements $Q_i$.  In fact, they give us various ways of writing the $Q_i$;
for example, in the $N=3$ case, we find
$Q_3=\sigma Z_1\sigma^{-1}Z^{-1}_{3}$.  We have pictorial
representations for all the generators on the right-hand side of this
relation, so we may explicitly find the pictorial representation of
$Q_3$.  From Figure \ref{Q_3}, we see that $Q_3$ acts
only on the third strand while leaving the other two untouched.  For
clarity, we have indicated the twisting using arrows; one must start
form the top of the third strand and follow the arrows around all
faces of the cube.

This is the pictorial representation of $Q_3$.  By pulling the
strands tight, we find that this is precisely the operator which
leaves the strands entirely alone: the identity $\mathbb{1}$, namely, the trivial braid. 
This result is not unique to $Q_3$; we find that the
graphical representation for each of the $Q$s is simply the identity.

Although this cube representation is successful in describing the $T_i$, $Y_i$ and $Z_i$ generators of $\mathcal{D}_N\{Q\}$, it still only allows the $Q_i$ to be represented by trivial braids, and so is really only valid when $Q_i=\mathbb{1}$.  Therefore, this is simply a representation of $\mathcal{D}_N\{Q\}/\langle Q_i\rangle$, i.e.\ the elliptic braid group
\cite{JB,GPS} (see Figure \ref{F1}).   However, if we wish to allow for values of $Q_i$ {\em
  other} than unity, we need to modify our cube representation in some
way, which we now describe.

\subsubsection{The General Case $Q_i\ne\mathbb{1}$: Introducing Ribbons}\label{4.3.2}
To obtain a nontrivial pictorial representation which accommodates
$Q_i\ne\mathbb{1}$, we modify our cube representation by replacing the strands
by ribbons.  This modification is not unmotivated: in order to extend
the $\mathcal{A}_N\{Q\}$ representation to one for a $\mathcal{D}_N\{Q\}$, we increased the dimension of our space from two to three, and so it is reasonable to increase
the dimension of our strands.

Doing so is precisely what we need in order for our representation to
work for {\em all} $\mathcal{D}_N\{Q\}$s, not just those where the $Q_i=\mathbb{1}$.  We therefore no
longer braid one-dimensional strands, but do so instead with
two-dimensional {\em ribbons}.  This extra degree of freedom will
enable us to completely describe a double affine $Q$-dependent braid group for any $Q_i$.

\begin{center}\begin{figure}[h!]
\includegraphics[scale=0.28]{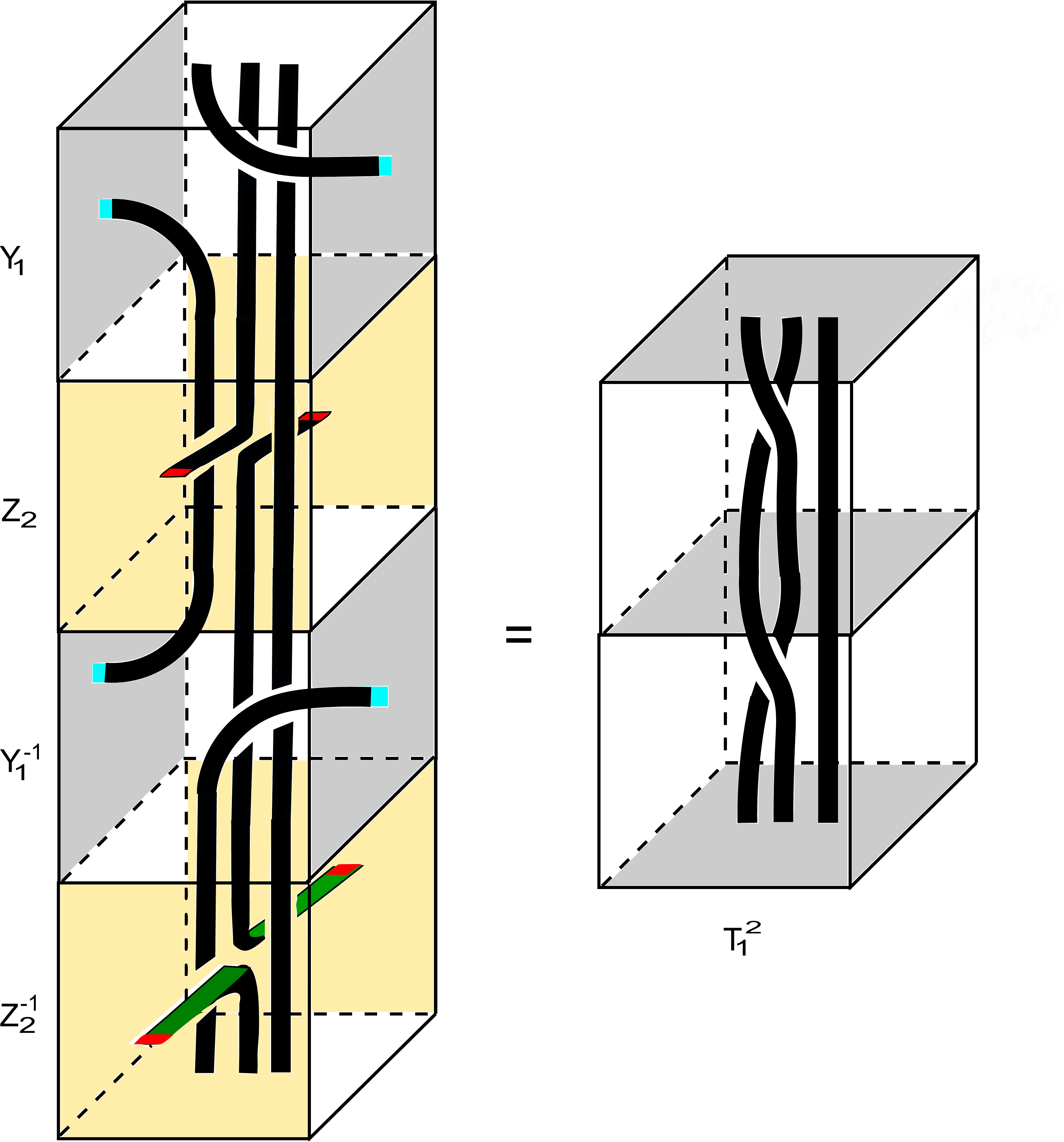}
\caption{The relation $Y_1Z_2Y_1^{-1}Z_2^{-1}=T_1^2$ using ribbons instead
of strands. Note the colour conventions.}
\label{T_1^2=Y_1Z_2Y_1Z_2ribbon}
\end{figure}\end{center}

However, before we revisit the elements $Q_i$, we must verify
that all of the previous $\mathcal{D}_N\{Q\}$ axioms still hold when using ribbons
within our cube representation.  It is straightforward to show that
they do; to illustrate this point, we explicitly show (\ref{YZT}), as
this relation contains all three types of generators, the $T_i$, $Y_i$
and $Z_i$.  (For clarity, we have coloured the front and back of each
ribbon respectively by black and green.)  This example, illustrated in Figure \ref{T_1^2=Y_1Z_2Y_1Z_2ribbon}, also allows us
to clearly lay out the braiding conventions that we use.

When the ribbon wraps in a left/right direction -- representing a
$Y_i$ operator -- we use turquoise for the tips that are identified
with each other.  It is vital to stress that these link the left and
right faces of the cube in a very particular fashion: the ribbon must
pass through a left or right face of the cube oriented {\em
  vertically}.  This condition ensures that the ribbon doesn't twist
while wrapping around the cube.

In a similar fashion, the ribbons representing the $Z_i$ generators
are coloured so that when a red tip is visible, this implies that the
ribbon passes through either the back or front face of the cube.  We
require that whenever such a ribbon intersects the front or back face
of the cube, it does so oriented {\em horizontally}.

These conventions give Figure \ref{T_1^2=Y_1Z_2Y_1Z_2ribbon} for $Y_1Z_2Y^{-1}_1Z^{-1}_2 =
T^2_1$, and pulling the ribbons tight we can clearly see that the relation
holds.  All of the other relations are satisfied in a similar manner.

One of the major advantages of our cube-ribbon representation is that specific crossing rules are not required when one ribbon crosses another. This is due to the fact that, following the conventions outlined above, the ribbons can braid in three distinct orthogonal directions and hence no such rules are necessary. In contrast, for framed braids in an infinitely long strip as in \cite{KO} more complicated crossing conditions are needed.

We now revisit the relation $Q_3=\sigma
Z_1\sigma^{-1}Z^{-1}_{3}$ which, when represented by 1-dimensional
strands, was equivalent to the identity element.  Now using ribbons
instead of strands, we construct the pictorial representation of
$Q_3$.  (For clarity, we show only the third ribbon, as this
is the only one which behaves nontrivially.)  Keeping with the colour
convention defined earlier, we obtain $Q_3$, and, by pulling
the ribbons tight, yields the key result we require: a twist in the
ribbon is created!  This important result is illustrated in Figure \ref{twistq}
below.

\begin{center}\begin{figure}[h!]
\includegraphics[scale=0.28]{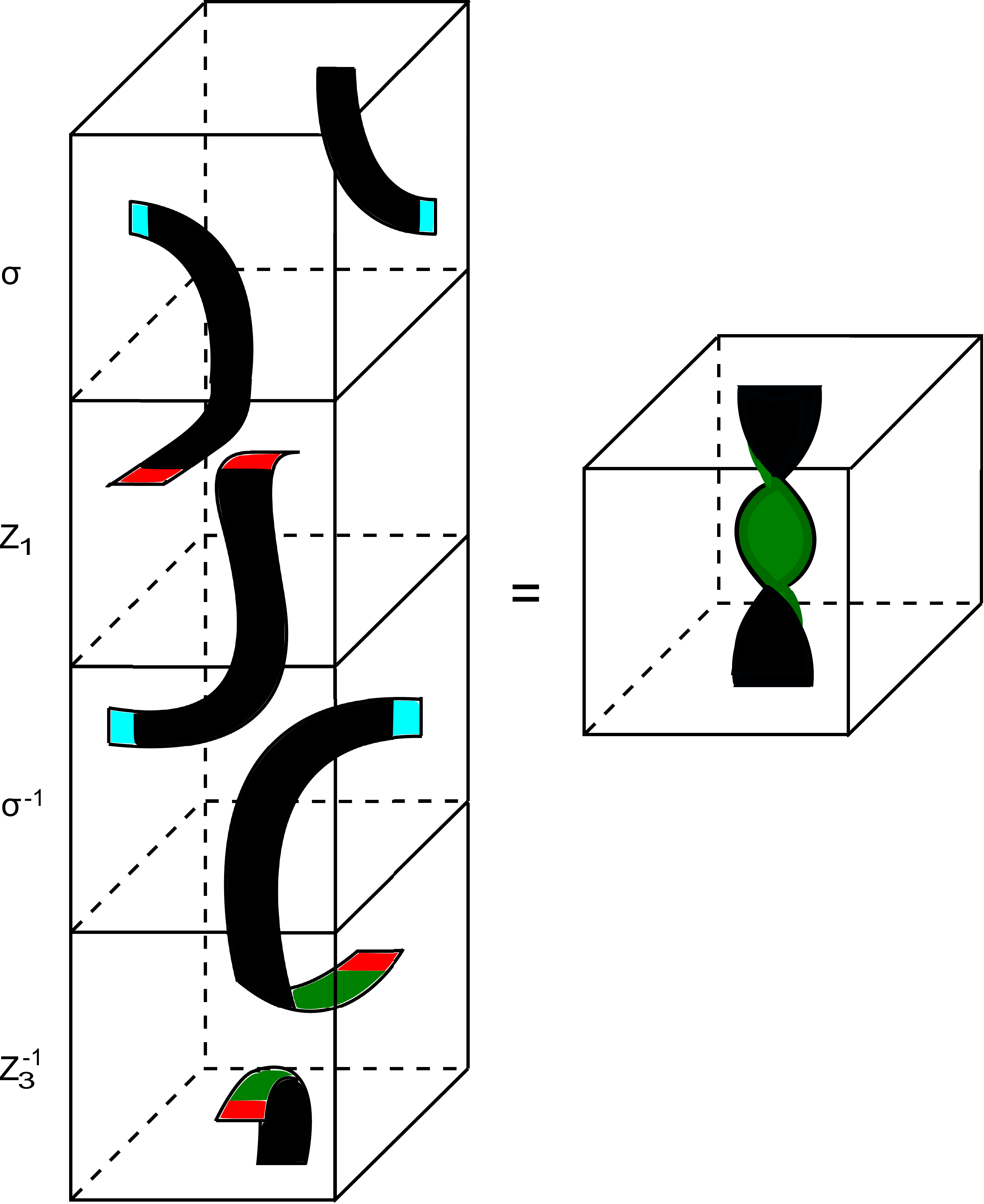}
\caption{$Q_3$, the creation of a twist in the third ribbon. The full anticlockwise twist
makes clearly visible both front and back faces of the ribbon, coloured black and green respectively. Note that we illustrate only the third ribbon.}
\label{twistq}
\end{figure}\end{center}

As this is the most important feature of our ribbon representation,
let us explain in detail how this comes about: in constructing $\sigma
Z_1\sigma^{-1}Z^{-1}_{3}$, both the black and green faces of the
ribbon are clearly visible.  Upon closer inspection, we see that the
ribbon undergoes a {\em full anticlockwise twist} in going from the
top face to the bottom one.  First, the front black face of the ribbon
is visible.  Then, having undergone half an anticlockwise twist, the
back green face becomes visible until finally the full anticlockwise
twist leaves the black face facing forwards.

This significant result can be generalised.  We have just shown that in our cube-ribbon representation $Q_3$ creates a twist in the third ribbon.  It is easily shown, following the construction of $Q_3$, that in our particular representation the action of $Q_i$ is to create a single full anticlockwise twist in the $i^{th}$ ribbon.

As the creation of a full anticlockwise twist in the ribbon may be somewhat
difficult to visualise we have included a more rigorous argument
to convince the reader in Appendix B.

Other expressions could be used to determine $Q_i$; for example,
(\ref{YZ}) gives
\begin{eqnarray*}
Q_i&=&Y_i\left(\prod_{j=1}^{N}Z_j\right)Y^{-1}_i\left(\prod_{j=1}^{N}
Z^{-1}_j\right).
\end{eqnarray*}
Or we could use (\ref{Z-sigma}):
$Q_i=\sigma^{N}Z_i\sigma^{-N}Z^{-1}_i$.  For these and any
other representation for $Q_i$, the result is the same,
namely, $Q_i$ creates a single full anticlockwise twist in the $i^{th}$ ribbon.

We can also verify that an expression like $Z_3\sigma
Z_1^{-1}\sigma^{-1}$, which the $\mathcal{D}_N\{Q\}$ axioms require to be
$Q^{-1}_3$ for $N=3$, is indeed a full {\em clockwise} twist
in the third ribbon, again totally consistent with our interpretation
of $Q_i$.
\smallskip

The interpretation of $Q_i$ is now clear: it is the generator that creates a full
anticlockwise twist in the $i^{th}$ ribbon.  Similarly $Q^{-1}_i$ creates a full 
clockwise twist in the $i^{th}$ ribbon.  As these are no longer trivial actions on the ribbons, we have a pictorial representation for $Q_i\neq \mathbb{1}$, and a full description for  $\mathcal{D}_N\{Q\}$.

\section{Double Affine Hecke Algebras}
\setcounter{equation}{0}
\renewcommand{\theequation}{\thesection.\arabic{equation}}

In the previous section we highlighted the fact that the elliptic braid group is given by $\mathcal{D}_N\{Q\}$/$\langle Q_i\rangle$.  Similarly readers familiar with double affine Hecke algebras \cite{IC1, BGHP} may recognise that our definition of a $\mathcal{D}_N\{Q\}$ closely resembles that of a double affine Hecke algebra (DAHA) without the Hecke relation.  We will in fact show precisely how to obtain a DAHA given our construction of a double affine $Q$-dependent braid group $\mathcal{D}_N\{Q\}$. 

\subsection{The Double Affine Hecke Algebra within $\mathcal{D}_N\{Q\}$ }

Consider the subgroup $\mathcal{C}$ of the $Q$-dependent braid group $\mathcal{B}_{N}\{Q\}$ defined as
\begin{eqnarray*}
\mathcal{C}&=&\langle Q_iQ^{-1}_{i+1},i=1,\ldots,N-1\rangle.
\end{eqnarray*}
It can easily be shown that $\mathcal{C}$ is a normal subgroup of $\mathcal{B}_{N}\{Q\}$, and so we can construct the quotient $\mathcal{G}=\mathcal{B}_{N}\{Q\}/\langle\mathcal{C}\rangle$, which is precisely the group we require to define a DAHA.  Within $\mathcal{G}$, the $Q_i$ are indistinguishable from one another; therefore, we refer to each of their cosets $[Q_i]$ as $Q$.  Most importantly, using (\ref{T3})-(\ref{T6}), we see that $Q$ now commutes with not only the squares of the braid group generators $T^2_i$, but also with the $T_i$ themselves.  We are now in a position to extend the quotient group $\mathcal{G}$ to a Hecke algebra.

\subsection{The Hecke Algebra $\mathcal{H}_{N}(t)$}

Before defining a DAHA, we must extend our quotient group $\mathcal{G}$ to an algebra in which the $T_i$ generators satisfy a particular relation; this defines the Hecke algebra.

Associate with $\mathcal{G}$ the Hecke algebra
$\mathcal{H}_{N}(t)$.  This is the group algebra of $\mathcal{G}$
over a field $k$ parametrised by $t\in k$ such that each generator
$T_i$ satisfies the {\it Hecke relation}
\begin{eqnarray}
\left(T_{i}-t^{1/2}\mathbb{1}\right)\left(T_i+t^{-1/2}\mathbb{1}\right)&=&0.
\label{Hecke}
\end{eqnarray}
It is worth noting that even though $T^{-1}_i$ was assumed to exist in
$\mathcal{G}$, this relation gives its form explicitly:
\begin{eqnarray*}
T^{-1}_i& =& T_i - \left(t^{1/2}-t^{-1/2}\right)\mathbb{1}.
\end{eqnarray*}

\subsection{The Double Affine Hecke Algebra $\mathcal{D}_{N}(t,q)$}

To complete the DAHA construction we must firstly extend the Hecke algebra $\mathcal{H}_{N}(t)$ to an Affine Hecke Algebra $\mathcal{A}_{N}(t)$.  This is achieved with the introduction of $N$ invertible operators $Y_i$
which satisfy (\ref{Y1})-(\ref{Y3}).

Recall that the $\mathcal{A}_{N}$ was fully generated by $Y_1$ and the $T_i$.  
It is perhaps worth pointing out that the affine Hecke algebra is also fully generated by $Y_1$ and the $T_i$, and we can reorder them as necessary. This was not true for the $\mathcal{A}_N\{Q\}$ as we need the full Hecke algebraic structure in order to consistently order the operators.  For
example, $T_1$ and $Y_3$ can be reordered as we like, but this is true
for $T_1$ and $Y_2$ only if we invoke the Hecke relation:
\begin{eqnarray*}
T_1Y_2&=&Y_2T_1^{-1}\\
&=&Y_2\left[T_1-\left(t^{1/2}-t^{-1/2}\right)\mathbb{1}\right]\\
&=&Y_2T_1-\left(t^{1/2}-t^{-1/2}\right)Y_2.
\end{eqnarray*}

Following \cite{IC1,MK} we take a DAHA $\mathcal{D}_{N}(t,q)$ of type $A$ to be the algebra
generated by $T_i$, $Y_i$ and $Z_i$ which satisfy equations (\ref{T1})-(\ref{T2}), the Hecke relation (\ref{Hecke}) along with (\ref{Y1})-(\ref{Y3}) and (\ref{Z1})-(\ref{Z3}).  

In addition to these the $Y_i$ and $Z_i$ obey the intertwining relations \cite{IC1}
\begin{eqnarray}
Y_1Z_2Y^{-1}_1Z^{-1}_2 &=& T^2_1,\label{YZT1}\\
Y_i\left(\displaystyle\prod_{j=1}^N Z_j\right)
 &=&q\left(\displaystyle\prod_{j=1}^N Z_j\right)Y_i,\label{YZ1}\\
Z_i\left(\displaystyle\prod_{j=1}^NY_j\right) &=&
q^{-1}\left(\displaystyle\prod_{j=1}^N Y_j\right)Z_i,\label{ZY1}
\end{eqnarray} 
where $q\in k$.

(As in the $\mathcal{D}_N\{Q\}$ (\ref{ZY1}) is not independent of the other relations, although it is often included in the literature as part of the definition of a DAHA.)

One must note that unlike our definition of the $\mathcal{D}_N\{Q\}$ where we have a set of elements $Q_i$, in the DAHA $q$ is simply a parameter.  So a DAHA  $\mathcal{D}_{N}(t,q)$ depends on the two variables $t,q$. This is entirely consistent with our construction of a DAHA from $\mathcal{D}_N\{Q\}$ via the quotient group $\mathcal{G}$ if we set $Q=q\mathbb{1}$.  We therefore have a representation of a DAHA in $\mathcal{B}_{N}\{Q\}/\langle\mathcal{C}\rangle$ when we impose $Q=q\mathbb{1}$.

In terms of the cube representation we can replace a ribbon with a full anticlockwise 
twist by one with no twist at all, only if we multiply the resulting cube by a factor 
of $q$.  One may see this explicitly in Figure \ref{twistq}.  As a result, one may view this twist as the first Reidemeister move on a ribbon:
\begin{center}
\includegraphics[scale=0.26]{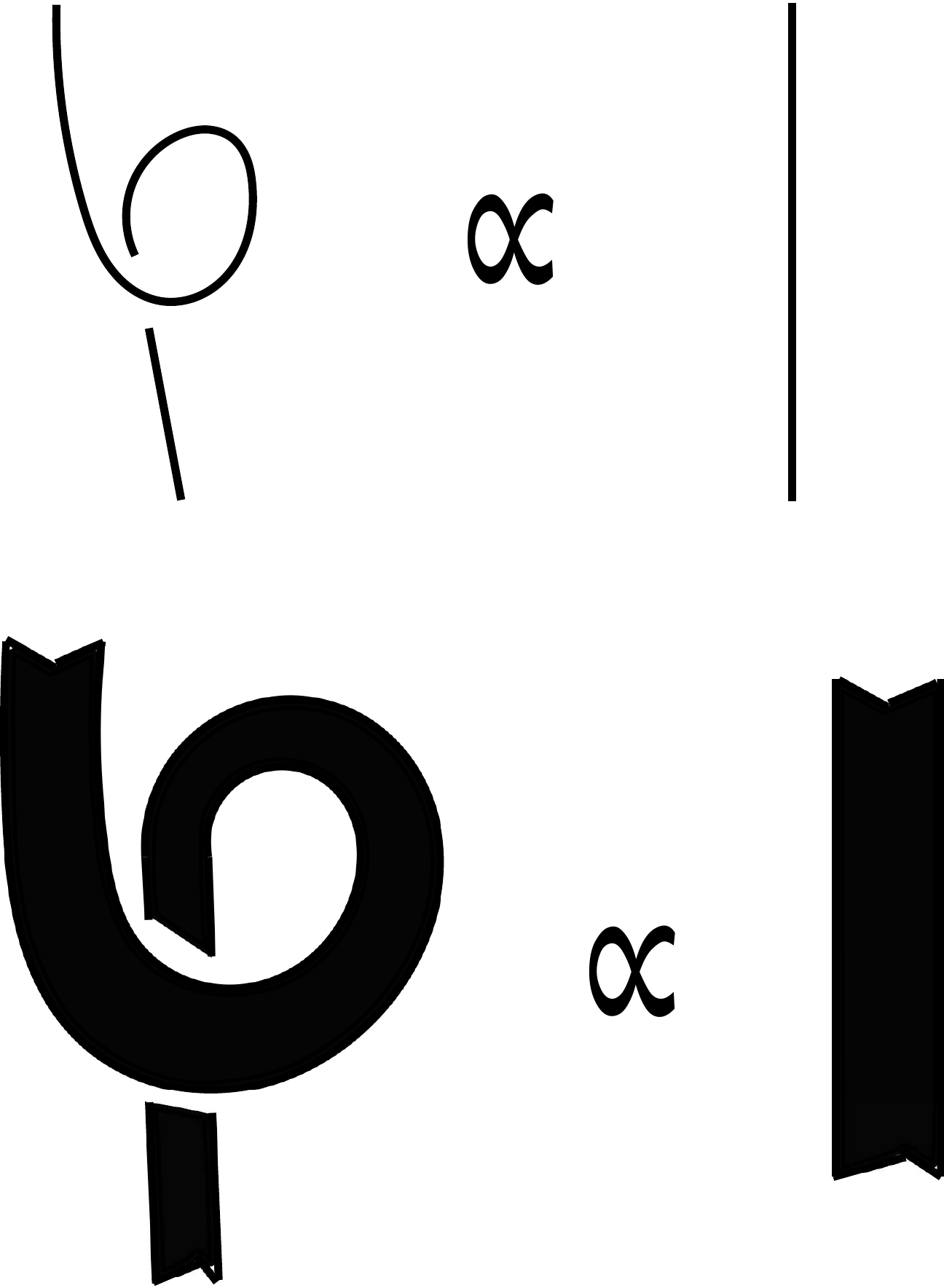}
\end{center}
Therefore the interpretation of $q$ is clear: it is the multiplicative factor in front of a DAHA element whenever we replace a ribbon with a full anticlockwise twist by one with no twist at all.
Furthermore since $q$ does not describe the actual position of the twist in the ribbon, one can have a factor of $q^n$ in front of a DAHA element corresponding to $n$ anticlockwise twists occurring {\em anywhere} in the cube.
As there is no restriction on what value $q$ can take, we are not limited to the case $q$=1 and  have a pictorial representation that fully describes any DAHA.

\section{Summary and Discussion}
\setcounter{equation}{0}
\renewcommand{\theequation}{\thesection.\arabic{equation}}

In this paper we have defined and presented a graphical representation of the double affine $Q$-dependent braid group.  Following the method of extending the pictorial representation
of the $Q$-dependent braid group to one for an $\mathcal{A}_N\{Q\}$, we found that all of the
relations not explicitly involving the operators $Q_i$ could be
satisfied by a $\mathcal{D}_N\{Q\}$ depicted using $1$-dimensional strands embedded in
a cube whose opposing vertical sides were identified, i.e.\ a
hollowed-out toroid.

This representation was consistent only for a $\mathcal{D}_N\{Q\}$ where all the $Q_i=\mathbb{1}$; that is, the elliptic braid group.  However, by replacing the strands with ribbons, our cube representation allowed us
to capture all aspects of a $\mathcal{D}_N\{Q\}$ and gave us a
nice interpretation of the action of any $Q_i$ as a single full anticlockwise twist in the $i^{th}$ ribbons.  We thus obtain an intuitive pictorial representation
which clearly incorporates all of the structure of the more abstract
$\mathcal{D}_N\{Q\}$.

We showed that our new graphical representation is also valid for all DAHAs.  Our definition of a  $\mathcal{D}_N\{Q\}$ reduced to one of a double affine Hecke algebra simply by attaching the Hecke algebra to one of its quotient groups.  The DAHA  depends on two parameters $t$ and $q$.  We found that graphically, the parameter $q$ corresponds to a full anticlockwise twist in the ribbon.  

\smallskip

By construction, our representation should be closely related to
tangles and knot theory.  Using elementary tangles via Reidemeister
moves to describe this algebra appears quite possible; in fact, the
replacement of a full twist by a factor of $q$ is very much a
Reidemeister-like move.  This would indicate a relation between our cube-ribbon representation and elementary tangle representations of affine Hecke algebras; we
hope to look further into this suspected relationship.

Similarly, transforming this cube-ribbon representation to an
equivalent matrix representation is an interesting challenge.
We hope to use our new pictorial representation to bring this closer to reality.

\section*{Acknowledgments}

This work has been funded under the Irish Research Council Embark
Initiative Postgraduate scheme.  We would also like to acknowledge
funding from Science Foundation Ireland under the Principal
Investigator Award 10/IN.1/I3013.

\section*{Appendix A}
\setcounter{equation}{0}
\renewcommand{\theequation}{A.\arabic{equation}}

In this Appendix, we show that $\prod_{j=1}^NY_j=\sigma^N$.  Although
this identity is already well-known \cite{IC1}, we present a proof for
the interested reader.

\smallskip

Define the operator $P_k$ by
\begin{eqnarray}
P_k:&=&\sigma^k\left(T_1\ldots T_k\right)^{-1}\left(T_2\ldots
T_{k+1}\right)^{-1}\ldots\left(T_{N-k}\ldots T_{N-1}\right)^{-1}.\label{P}
\end{eqnarray}
We want to show by induction that this is equal to
$P_k=\displaystyle\prod_{j=N-k+1}^NY_j$.
\begin{enumerate}
\item For $k=1$:
\begin{eqnarray*}
P_1&:=&\sigma^1 \left(T_1\right)^{-1}\left(T_2\right)^{-1}\ldots\left(
T_{N-1}\right)^{-1}\\
&=&\sigma T^{-1}_1T^{-1}_2\ldots T^{-1}_{N-1}\\
&=&Y_N,
\end{eqnarray*}
so $P_1$ is indeed equal to $\displaystyle\prod_{j=N-1+1}^NY_j=Y_N$,
and the assertion is true for $k=1$.
\item Now assume that our assertion is true for some $k$, namely,
\begin{eqnarray*}
P_k&=&\sigma^k\left(T_1\ldots T_k\right)^{-1}\left(T_2\ldots
T_{k+1}\right)^{-1}\ldots\left(T_{N-k}\ldots
T_{N-1}\right)^{-1}=\prod_{j=N-k+1}^NY_j.
\end{eqnarray*}
If this holds, then $P_kY_{N-k}$ is $\displaystyle\prod_{j=N-k}^NY_j$
because all the $Y_i$ commute.  Using $Y_{N-k}=T_{N-k}\ldots
T_{N-1}\sigma T^{-1}_1\ldots T^{-1}_{N-k-1}$, we can rewrite this same
expression as
\begin{eqnarray*}
P_kY_{N-k}&=&\left[\sigma^k\left(T_1\ldots T_k\right)^{-1}\left(T_2\ldots T_{k+1}
\right)^{-1}\ldots\left(T_{N-k}\ldots T_{N-1}\right)^{-1}\right]\\
&&\times
\left[T_{N-k}\ldots T_{N-1}\sigma T^{-1}_1\ldots T^{-1}_{N-k-1}\right]\\
&=&\left[\sigma^k\left(T_1\ldots T_k\right)^{-1}\left(T_2\ldots T_{k+1}
\right)^{-1}\ldots\left(T_{N-k-1}\ldots T_{N-2}\right)^{-1}\right]\\
&&\times
\left[\sigma T^{-1}_1\ldots T^{-1}_{N-k-1}\right].
\end{eqnarray*}
Using $T^{-1}_i\sigma=\sigma T^{-1}_{i+1}$, all $\sigma$s can be moved
to the left:
\begin{eqnarray*}
P_kY_{N-k}&=&\left[\sigma^{k+1}\left(T_2\ldots T_{k+1}\right)^{-1}
\left(T_3\ldots T_{k+2}\right)^{-1}\right]\ldots\\
&&\dots
\left[\left(T_{N-k}\ldots T_{N-1}\right)^{-1}
T^{-1}_1\ldots T^{-1}_{N-k-1}\right].
\end{eqnarray*}
$T_i$ commutes with all other $T$s except $T_{i+1}$ and $T_{i-1}$, so
we may pull the rightmost operators $T_1^{-1}$ to $T_{N-k-1}^{-1}$ as
far as possible to the left:
\begin{eqnarray*}
P_kY_{N-k}&=&\sigma^{k+1}\left[\left(T_2\ldots T_{k+1}\right)^{-1}T_1^{-1}\right]
\left[\left(T_3\ldots T_{k+2}\right)^{-1}T_2^{-1}\right]\ldots\\
&&\dots
\left[\left(T_{N-k}\ldots T_{N-1}\right)^{-1}T^{-1}_{N-k-1}\right]\\
&=&\sigma^{k+1}\left(T_1\ldots T_{k+1}\right)^{-1}
\left(T_2\ldots T_{k+2}\right)^{-1}\ldots
\left(T_{N-k-1}\ldots T_{N-1}\right)^{-1}.
\end{eqnarray*}
But (\ref{P}) tells us that this is precisely the definition of
$P_{k+1}$.  Thus, $P_kY_{N-k}=P_{k+1}$, so
$P_{k+1}=\displaystyle\prod_{j=N-k}^NY_j$ and our assertion holds for
$k+1$ if it holds for $k$.
\end{enumerate}
This therefore verifies that
\begin{eqnarray*}
\sigma^k\left(T_1\ldots T_k\right)^{-1}\left(T_2\ldots
T_{k+1}\right)^{-1}\ldots\left(T_{N-k}\ldots T_{N-1}\right)^{-1}
&=&\prod_{j=N-k+1}^NY_j
\end{eqnarray*}
for all $k=1,2,\ldots,N-1$.  For $k=N-1$, this gives
\begin{eqnarray*}
\sigma^{N-1}\left(T_1\ldots T_{N-1}\right)^{-1}&=&\prod_{j=2}^NY_j.
\end{eqnarray*}
But $\sigma^{-1}(T_1\ldots T_{N-1})^{-1}=Y_1^{-1}$, so we find that
\begin{eqnarray*}
\prod_{j=1}^NY_j=\sigma^N.&&\blacksquare
\end{eqnarray*}

\section*{Appendix B}
\setcounter{equation}{0}
\renewcommand{\theequation}{B.\arabic{equation}}

Here we show that the twist in the ribbon generated by $Q_3$ is precisely $2\pi$.  We demonstrate this specifically for the case of $Q_3=\sigma Z_1\sigma^{-1}Z^{-1}_3$ as in Figure (\ref{twistq}) where, from top to bottom, a full anticlockwise twist in the third ribbon is obtained. For clarity we illustrate only the third ribbon as it is the only one that behaves non-trivially.
  
Firstly let $z(s), (0\le s \le 1)$ denote the position of a point on the ribbon.  Then $\hat{v}$ is the unit vector indicating the ribbon orientation and always lies on the surface of the ribbon.  The direction of motion is given by the unit vector $\hat{u}$, where at all times $\hat{u}.\hat{v}=0$.  The vector $\hat{w}=\hat{u}\times\hat{v}$ defines the normal to the ribbon.\\
So there is an orthogonal frame $g(s)=[\hat{u},\hat{v},\hat{w}]$ attached to each point on the ribbon as indicated in the diagram below.

\begin{center}
\includegraphics[scale=0.28]{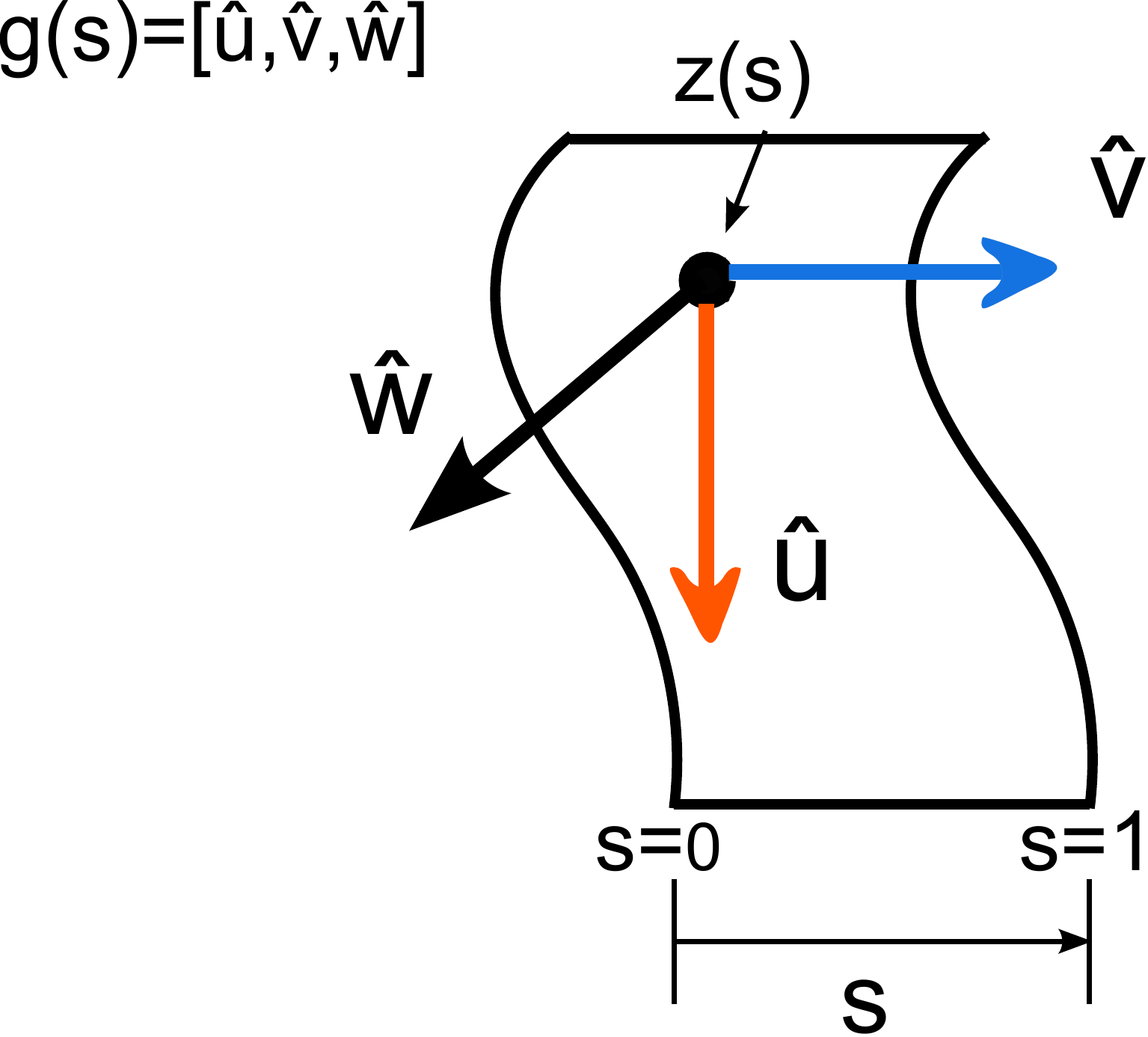}
\end{center}

We now follow a point as it travels down the ribbon.  Attached to this point is the orthogonal frame $g(s)$. We impose that the ribbon cannot twist around the direction of motion, that is; $\omega.\hat{u}=0$ where $\omega$ is the angular velocity of the frame $g(s)$.  We measure the degree of rotation of $g(s)$, between the top and bottom of the ribbon, relative to a fixed frame.  This yields the size of the twist in the ribbon.

\begin{center}\begin{figure}[h!]
\includegraphics[scale=0.26]{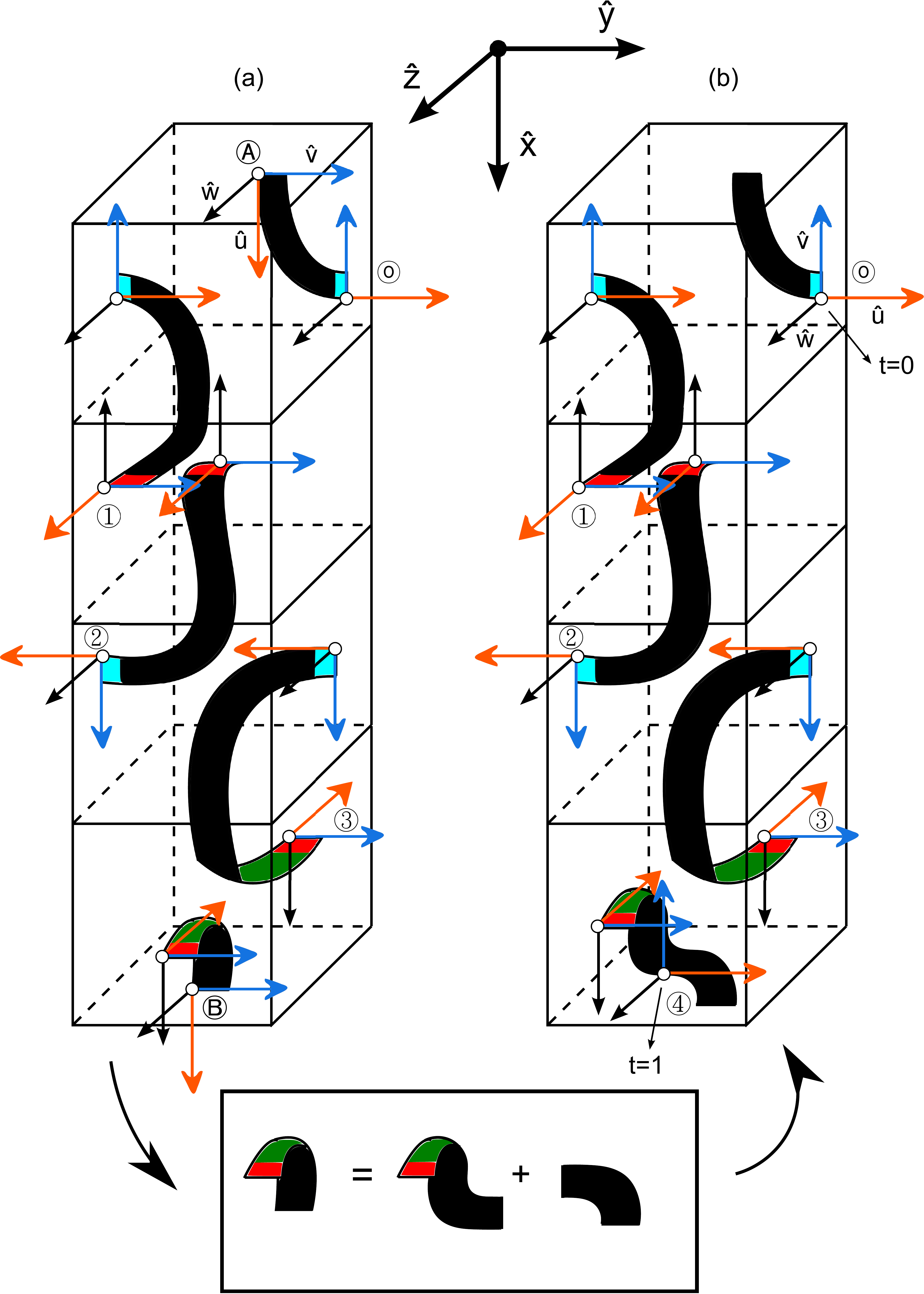}
\caption{Figure (a) shows $g(s)$ at various points along the ribbon $Q_3=\sigma Z_1\sigma^{-1}Z^{-1}_3$. In Figure (b) we redraw the relation such that between times $t=0$ and $t=1$ one can see $\hat{u}$ rotating by $2\pi$ in the $\hat{y}-\hat{z}$ plane.}
\label{rotation}
\end{figure}\end{center}

The Figure \ref{rotation} (a) below shows the frame $g(s)$ at various points along the ribbon, from the top of the ribbon labelled point (A), to the bottom of the ribbon; point (B).  Between these points we show that the moving frame $g(s)$ undergoes a full $2\pi$ rotation relative to the inertial reference frame ($\hat{x},\hat{y},\hat{z}$).

Notice that between points (A) and (0), the ribbon itself does not undergo any rotation.  Therefore without losing any information we can measure the twist starting from point (0), which we now call time $t=0$, as in Figure \ref{rotation} (b).\\
Furthermore in Figure \ref{rotation} (b), the bottom of the ribbon is redrawn in such a way that the extra turns do not contribute to the overall twist.  Then following $g(s)$ from $t=0$ to $t=1$, one can immediately see that $\hat{u}$ rotates {\em only} in the $\hat{y}-\hat{z}$ plane.  In fact it does exactly a $2\pi$ clockwise rotation.  So at any time $t$,  $\hat{u}$ can be written as follows:
\begin{center}
$\hat{u}(t)=\cos(2\pi t)\hat{y}+\sin(2\pi t)\hat{z}$.
\end{center}
One can easily check this holds.  For example at time $t=1/2$, $\hat{u}(1/2)=-\hat{y}$.  This is
verified upon inspection of point (2) in the diagram.  

Further inspection reveals that as $\hat{u}$ rotates in the $\hat{y}-\hat{z}$ plane, the vectors $\hat{v}$ and $\hat{w}$ rotate in a clockwise fashion around $\hat{u}$.\\
We introduce a frame $[\hat{e}_1,\hat{e}_2,\hat{e}_3]$, where $\hat{e}_1=\hat{u}$ and $\hat{e}_2$, $\hat{e}_3$ are functions of $\hat{v}$ and $\hat{w}$, to measure the rotation of $\hat{v}$ and $\hat{w}$ around $\hat{u}$.  Impose that at $t=0$, $\hat{e}_1=\hat{u},\hat{e}_2=\hat{v}$ and $\hat{e}_3=\hat{w}$.  It is important to note that $\hat{e}_1=\hat{u}$ at all times; that is we have $\hat{u}(t)=\hat{e}_1$.\\
Therefore in terms of this frame $[\hat{e}_1,\hat{e}_2,\hat{e}_3]$ we can write:
\begin{center}
$\hat{v}(t)=\cos(2\pi t)\hat{e}_2-\sin(2\pi t)\hat{e}_3$,\\
$\hat{w}(t)=\sin(2\pi t)\hat{e}_2+\cos(2\pi t)\hat{e}_3$.
\end{center}

Again these can easily be verified through simple substitution and by referring to the above diagram.  

$\hat{u}$ was fixed to $\hat{e}_1$ so in terms of the inertial reference frame we have:
\begin{center}
$\hat{e}_1(t)=\cos(2\pi t)\hat{y}+\sin(2\pi t)\hat{z}$.
\end{center}

Following the vector $\hat{e}_2$ between $t=0$ and $t=1$ we see that it always points in the negative $\hat{x}$ direction. This implies that:
\begin{center}
$\hat{e}_2(t)=-\hat{x}$.
\end{center}

Since $[\hat{e}_1,\hat{e}_2,\hat{e}_3]$ form an orthogonal frame we must have that:
\begin{center}
$\hat{e}_3(t)=-\sin(2\pi t)\hat{y}+\cos(2\pi t)\hat{z}$.
\end{center}

Finally in terms of the fixed frame ($\hat{x},\hat{y},\hat{z}$);
\begin{eqnarray*}
\hat{u}(t)&=&\cos(2\pi t)\hat{y}+\sin(2\pi t)\hat{z},\\
\hat{v}(t)&=&-\cos(2\pi t)\hat{x}+{\sin}^2(2\pi t)\hat{y}-\sin(2\pi t)\cos(2\pi t)\hat{z},\\
\hat{w}(t)&=&-\sin(2\pi t)\hat{x}-\sin(2\pi t)\cos(2\pi t)\hat{y}+{\cos}^2(2\pi t)\hat{z}.
\end{eqnarray*}

One can clearly see that $\hat{v}$ undergoes a full $2\pi$ clockwise rotation from $t=0$ to $t=1$.  $\hat{v}$ lies on the ribbon surface at all times, therefore requiring the ribbon to undergo the same rotation.  This yields precisely the required result; $Q_3$ creates a full anticlockwise twist in the third ribbon.

\end{document}